\documentclass[times,twocolumn,final]{elsarticle}

\usepackage{medima}
\usepackage{framed,multirow}

\usepackage[table]{xcolor}
\usepackage[disable]{todonotes}
\usepackage{soul}

\renewcommand\hl[1]{#1} 

\usepackage{float}
\usepackage{subcaption}
\usepackage{algorithmic}
\usepackage{graphicx}
\usepackage{hyperref}
\usepackage{comment}
\usepackage{array}
\usepackage{wrapfig}
\usepackage{amsmath}
\usepackage{subcaption}
\usepackage[scaled]{helvet}
\usepackage{placeins}
\usepackage{textcomp}

\newsavebox{\threesubbox}
\usepackage{dblfloatfix}
\usepackage{stackengine}
\usepackage{nicematrix}

\newcommand\Tstrut{\rule{0pt}{2.6ex}}         
\newcommand\Bstrut{\rule[-0.9ex]{0pt}{0pt}}   
\usepackage{amssymb}
\usepackage{latexsym}

\usepackage{url}
\usepackage{setspace}

\makeatletter

\renewcommand\p@subfigure{}
\makeatother
\DeclareCaptionLabelFormat{prefixsubfig}{Fig#1. #2 }
\usepackage{geometry}

\definecolor{newcolor}{rgb}{.8,.349,.1}

\begin{document}

\verso{Sophie Ostmeier \textit{et~al.}}

\begin{frontmatter}

\title{USE-Evaluator: Performance Metrics for Medical Image Segmentation Models Supervised by Uncertain, Small or Empty Reference Annotations in Neuroimaging}

\author[1]{Sophie \snm{Ostmeier}\corref{cor1}}
\cortext[cor1]{Corresponding author:}
  \ead{sostm@stanford.edu}
\author[1]{Brian \snm{Axelrod}}
\author[2]{Fabian \snm{Isensee}}
\author[3]{Jeroen \snm{Bertels}}
\author[1]{Michael \snm{Mlynash}}
\author[4]{Soren \snm{Christensen}}
\author[1]{Maarten G. \snm{Lansberg}}
\author[1]{Gregory W. \snm{Albers}}
\author[5]{Rajen \snm{Sheth}}
\author[3]{Benjamin F.J. \snm{Verhaaren}}
\author[1]{Abdelkader \snm{Mahammedi}}
\author[1]{Li-Jia \snm{Li}}
\author[1]{Greg \snm{Zaharchuk}}
\author[1]{Jeremy J. \snm{Heit}}

\address[1]{Stanford University, Center of Academic Medicine, 453 Quarry Rd, Palo Alto, CA 94304}
\address[2]{Division of Medical Image Computing, German Cancer Research Center (DKFZ), Im Neuenheimer Feld 280, 69120 Heidelberg, Germany}
\address[3]{KU Leuven, Herestraat 49, 3000 Leuven, Belgium}
\address[4]{gray number analytics, Lomma, Sweden}
\address[5]{in personal capacity}

\accepted{}

\begin{abstract}
Performance metrics for medical image segmentation models are used to measure the agreement between the reference annotation and the predicted segmentation. 
Usually, overlap metrics, such as the Dice, are used as a metric to evaluate the performance of these models in order for results to be comparable. 

However, there is a mismatch between the distributions of cases and the difficulty level of segmentation tasks in public data sets compared to clinical practice.  
Common metrics used to assess performance fail to capture the impact of this mismatch, particularly when dealing with datasets in clinical settings that involve challenging segmentation tasks, pathologies with low signal, and reference annotations that are uncertain, small, or empty. Limitations of common metrics may result in ineffective machine learning research in designing and optimizing models. To effectively evaluate the clinical value of such models, it is essential to consider factors such as the uncertainty associated with reference annotations, the ability to accurately measure performance regardless of the size of the reference annotation volume, and the classification of cases where reference annotations are empty.

We study how uncertain, small, and empty reference annotations influence the value of metrics on a stroke in-house data set regardless of the model.
We examine metrics behavior on the predictions of a standard deep learning framework in order to identify suitable metrics in such a setting. 
We compare our results to the BRATS 2019 and Spinal Cord public data sets.
We show how uncertain, small, or empty reference annotations require a rethinking of the evaluation.
The evaluation code was released to encourage further analysis of this topic https://github.com/SophieOstmeier/UncertainSmallEmpty.git
\end{abstract}


\end{frontmatter}

%
\FloatBarrier
\section{Introduction} 
\label{sec:introduction}

\begin{figure}
\centering
\begin{subfigure}[b]{0.40\textwidth}
\centering
\includegraphics[width=\textwidth]{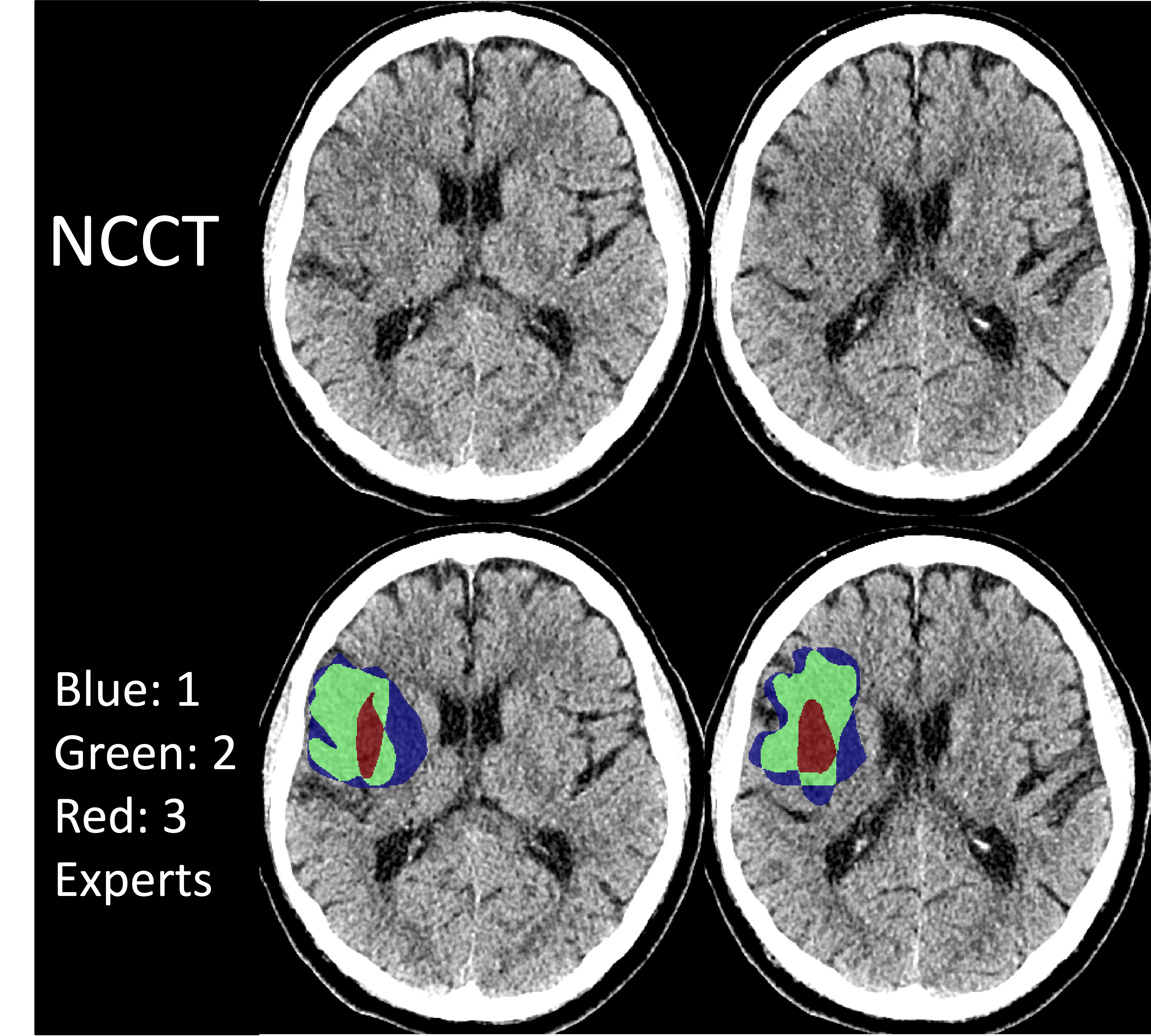}
\caption{Row 1: Uncertainty: Non-contrast Computed Tomography from an acute stroke patient within 16h. Row 2: Segmentation of all experts. The segmentations of all experts do not completely overlap. \label{fig:uncertainty}}
\end{subfigure}
\hfill
\begin{subfigure}[b]{0.40\textwidth}
\centering
\vspace*{\fill}
\includegraphics[width=\textwidth]{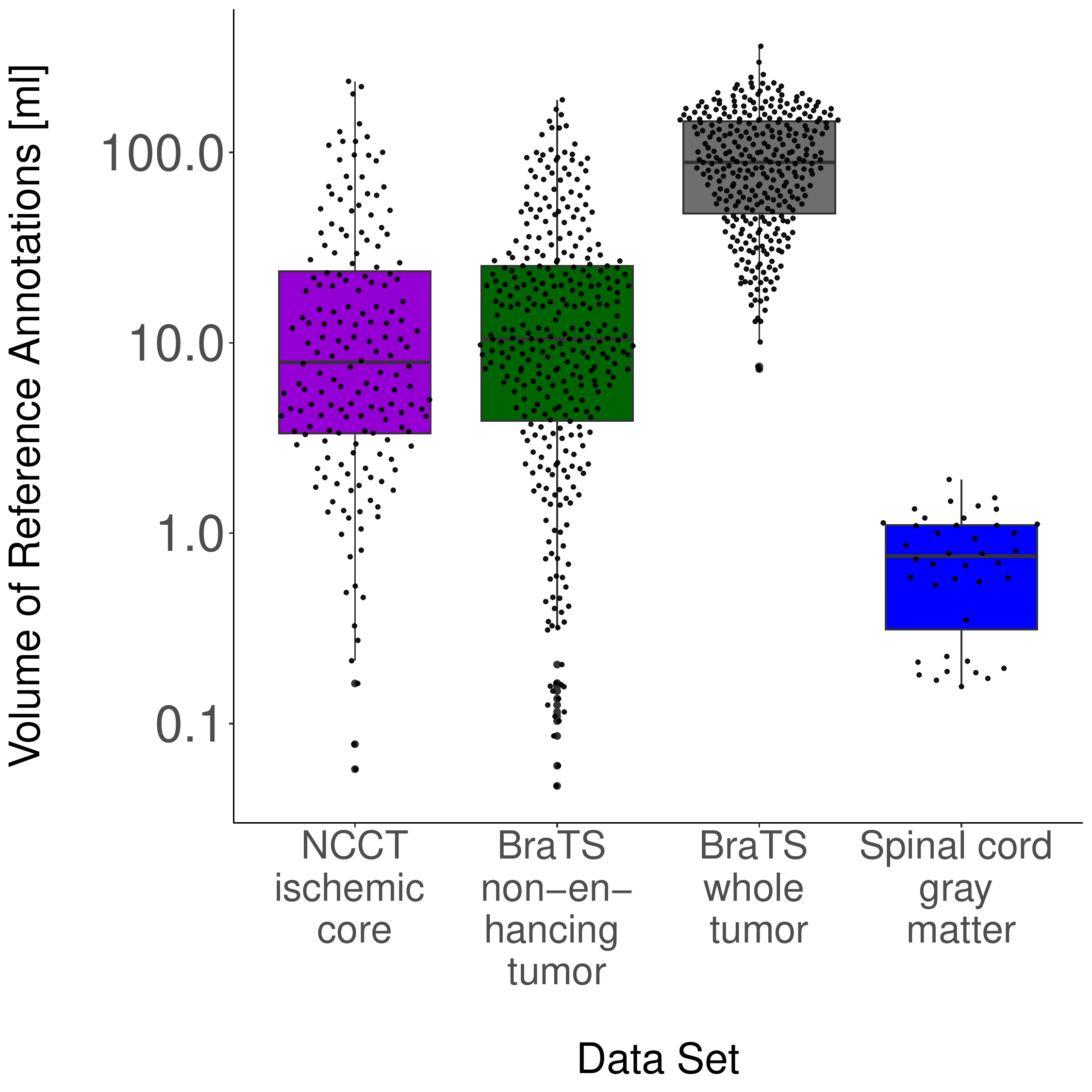}
\caption{\hl{Small Volumes: Boxplot with volume distribution of all data sets}\label{fig:distribution}}
\end{subfigure}
\hfill
\begin{subfigure}[b]{0.40\textwidth}
\centering
\includegraphics[width=\textwidth]{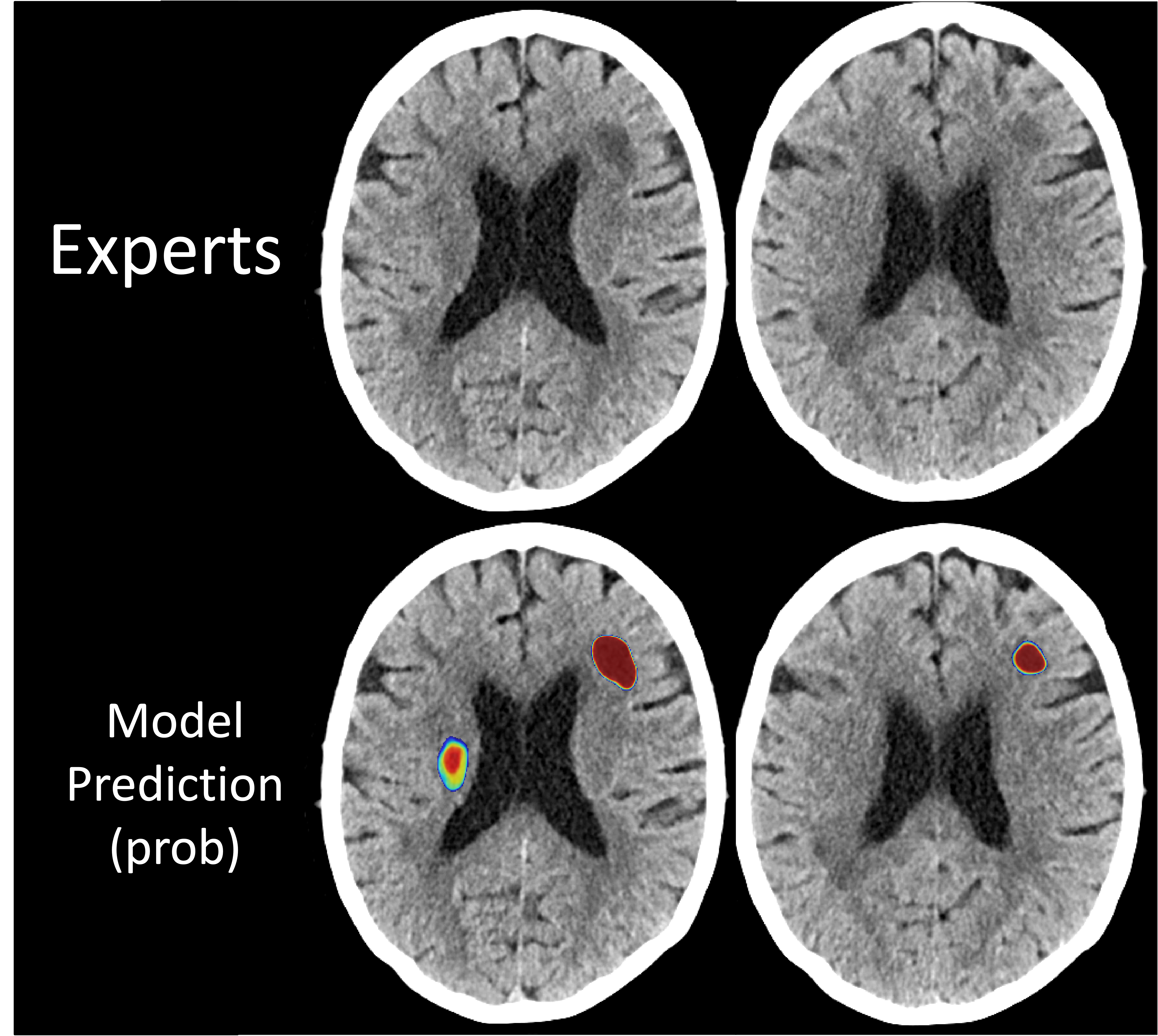}
\caption{Empty reference annotations: Row 1, Segmentation of all experts is “empty”. Row 2, predicted voxel probabilities (softmax output values) of the models (low to high probability indicated by blue to red colors)). All colored pixels may be "false positives".\label{fig:empty}}
\end{subfigure}
\end{figure}

The performance of machine learning algorithms is assessed by metrics. The optimal choice of metrics depends on the data set and the machine learning task to guarantee that the predictions accurately describe the intended phenomenon \citep{TahaMetric}.
Metrics can be used in two different ways. 
First, as the criteria that the models try to optimize as a loss function.
Second, as a way of validating and evaluating the performance of the model. 
This work focuses on the latter, referred to as performance metrics.

Performance metrics differ in their characteristics. The correlations between them determine the additional information revealed. Therefore, the appropriate selection of a performance metric for a specific task ensures consistency in model performance between development and deployment. 
For example, physicians that potentially use model predictions for treatment decisions of patients rely on an optimization and evaluation process of the models towards reliable and meaningful clinical information.

For data sets with uncertain (inter-expert variability), small ( e.g. $<1$\% of organ), or empty reference annotations, common metrics may penalize or misinterpret clinically meaningful information.
Prior studies have described the importance of quantifying uncertainty in the reference annotation \citep{mehta2022qu}, the dependency of metric values on the segmentation size and degree of class imbalance \citep{TahaMetric,LiuDWIstroke,CommowickMSSEG}, the equal weighting of all regions of misplaced delineation independently of their distance from the surface \citep{NikolovSDTheadneck} or missing definition for empty reference annotations \citep{CommowickMSSEG, Maier_Hein_pitfall_metric}. 

The failure to describe uncertain, small, or empty segmentations may lead to irrelevant and misleading optimization and evaluation procedures in segmentation models.

Here, we determine how to implement clinically meaningful metrics for medical segmentation models with the UncertainSmallEmpty \href{https://github.com/SophieOstmeier/UncertainSmallEmpty.git}{(USE)-Evaluator}.
We analyze the behavior of established metrics on benchmark deep learning models trained on four data sets with and without uncertain, small, and empty reference annotations (in-house and public).

\subsection{Uncertain Reference Annotations \label{sec:uncertainty} }

\hl{While experts annotations may exhibit variations in the volume and location of segmented objects, we assume that each expert possesses the highest level of human ability for the given task and, as a result, their judgments are considered equally valid }\citep{JungoUncertainty}. Identifying a superior expert who can definitively determine the correctness among the experts would require someone with even higher human ability. However, the process of appointing such an overruling expert would necessitate another individual with even greater abilities to make this judgment, leading to an infinite loop. \hl{Consequently, in the context of our study, we assume the absence of an overruling expert.}

For volume agreement, the reference annotation’s classification of a voxel can be true, and the segmentation of another expert or the prediction of the models can be false or vice versa. 
In practice, the spectrum ranges from a worst-case to a best-case scenario. In the best-case scenario, all false positives ($FP$) are truly positives. In the worst-case scenario, all $FP$ are truly $FP$. 
For example, in Figure \ref{fig:uncertainty} the union annotation of an acute stroke from experts A, B, and C (blue, green, and red) is larger than the majority vote (green and red). Some blue voxels at the border of the segmentation might falsely or truly be part of acute stroke ($FP$ or $TP$).
Another example is shown in Figure \ref{fig:empty}. Experts reference annotation is empty (first row). However, the prediction (second row) is not empty. Visual investigation shows an ambiguous lesion that was not segmented by the experts making all voxels $FP$ but maybe truly $TP$. 
The underlying low signal-to-noise ratio of stroke on Non-contrast Computed Tomography (NCCT) and the continuous transition from healthy to pathological brain tissue inevitably prevent a precise membership of these voxels. 

For location agreement, the distance between voxels from the reference annotation to another expert or prediction might be longer or shorter. For example, in Figure \ref{fig:uncertainty} the surface voxels of the green voxels will have a different distance than the blue surface voxels to the surface voxels of a predicted segmentation. 

In the BRATS 2019 data set, we reproduce an underlying low signal-to-noise ratio and a more continuous transition by using the non-enhancing tumor segmentation on native T1 as reference annotation. 
We compare to a high signal-to-noise ratio setting with a more discrete transition by using the whole tumor segmentation on T1, T1 enhanced, T2-flair, and T2 MRI images and Spinal Cord white matter segmentation on T2 MR images.

We propose the \hl{Uncertainty score (U-score)} as a quantifying measure of reference annotations uncertainty.


\subsection{Small Reference Annotations }

Depending on the clinical context, small reference annotations may be defined as relative to the total size of the studied body region. (e.g. less than 1\%). For the brain, 1\% is about 13 ml \citep{AkeretVolumeStandardization}. The distributions of reference annotation volumes vary across medical image data sets and segmentation tasks (Figure \ref{fig:distribution}) \citep{BakasBRATS19,BakasBRATS2019_2,MenzeBRATS19_3, prados2017spinal}. 

We hypothesize that the distribution of reference annotation volumes influences the value of metrics independently from the model’s performance (Figure \ref{fig:uncertainty})\citep{Maier_Hein_pitfall_metric}. For example, an acute ischemic stroke patient with a suspected large vessel occlusion undergoes emergent imaging to quantify the extent of the irreversible brain injury. The stroke volume is often quite small (1-5 ml in volume \citep{PowersManagement})). Models may segment a 1-2 ml lesion volume that has poor overlap with the segmentation by a neuroradiologist and have a low-performance metric despite properly identifying the volume. A slight difference in volume location within the brain is highly unlikely to influence a physician’s decision to treat the patient. 

We describe how the distribution of reference annotation volumes produces different metrics values, irrespective of the level of location and volume agreement between the model's predictions and the annotations.

\subsection{Empty Reference Annotations }
Images with empty reference annotations are described as masks in which the object of interest could not be identified by the annotators. The object might have been invisible at the time of the segmentation (Figure \ref{fig:empty}). 

Segmentation of an object within an image is a different task than a classification of an image. A classification task confirms the presence or absence of an object in the image (image-level), while a segmentation task assigns each voxel of the image to an object class (voxel-level) \citep{Maier_Hein_pitfall_metric}. 
An image-level classification task can also be formulated as a segmentation task by checking if the reference and predicted masks are empty. 
Therefore, when using a segmentation model in this way, it is important for the performance metrics to capture behavior on empty masks. 
\hl{However, some metrics for image segmentation return "NaN" or 0, if the model correctly predicts an empty mask (e.g. Dice, Specificity, Sensitivity, IoU).}

For clinical deployment, images with empty reference annotation are possible and their presence is crucial information. The predictions of segmentation models need to be optimized and evaluated for correct image-level classification \citep{CommowickMSSEG}. 
For example, it is possible that a stroke lesion in an early time window (0-4h after symptom onset) has a very low signal and cannot be segmented on NCCT. 
In this case, the reference annotation and the predicted segmentation should both be empty and an image-classification metric should return the optimal value. 
No visible and no predicted lesion would result in a treatment decision in favor of endovascular therapy \citep{PowersManagement}. 

We explore a potential solution by setting a volumetric threshold tailored to each clinical context where voxel-wise agreement is expected to go beyond clinical relevance. Below the threshold, the agreement between the reference annotation and prediction is automatically evaluated as an image-level classification task (e.g. stroke present or absent in the image \href{https://github.com/SophieOstmeier/UncertainSmallEmpty.git}{USE-Evaluator}). 
\subsection{Clinical Value}
\label{sec:clinical_value}

For a successful transition to \hl{clinical translatable challenge-winning segmentation} models, the focus on clinically meaningful optimization and performance metrics for each clinical context is crucial. Clinical value includes:
\begin{itemize} 
\item Robustness toward uncertainty in the reference annotation
\item Independence from the reference volume
\item Reward of volumetric and location agreement between the reference annotation and predictions
\item Evaluation of correct classification of empty reference annotations and predictions
\end{itemize}

\section{Metrics}

\subsection{Fundamentals }
\begin{table*}[t!]
\centering
\caption{Definitions of Performance Metrics for Medical Image Segmentation}
\label{table:metric_definition}
\begin{tabular}{|p{1.5cm}|p{4cm}|l|p{5cm}|l|}
\hline  
\textbf{Category}  & \textbf{Metric}   & \textbf{Abbr} & \textbf{Usage} & \textbf{Definition} \\ 
\hline 
\Tstrut \textbf{Volume}    & Volumetric Similarity    & VS   &   \citep{CaraduAorta,de_VosVessel,TiulpinOsteochondral,DeweyDeepHarmony,VaniaCNN}    & \Bstrut  \belowbaseline[0pt]{$1 - \frac{\left | \hat{V} -  V \right |}{  \hat{V} +  V + \epsilon}$} \\
& Absolute Volume Difference    & AVD  &   \citep{AmukotuwaCBF,BroschProstate}    &  $\frac 1 m \sum\limits_{j = 1}^m \left | V_j - \hat{V_j}\right |$   \\ \hline
\Tstrut\Bstrut \textbf{Overlap}   & Dice Similarity Coefficient   & Dice &   \citep{BeckerProstate,VaniaCNN,BroschProstate}    &  \belowbaseline[0pt]{$\frac{2\times TP}{2\times TP + FN + FP}$}\\
& Jaccard Index, Intersection over Union  & IoU &   \citep{bertels2019optimizing}   &  \abovebaseline[1pt]{$\frac{TP}{TP + FN + FP}$}\\
& Recall = Sensitivity& Recall    &   \citep{VaniaCNN}\Tstrut\Bstrut&   \abovebaseline[1pt]{$\frac{TP}{TP + FN}$}   \\
& Precision & Precision &   \citep{VaniaCNN}    &    \abovebaseline[0pt]{$\frac{TP}{TP + FP}$}\Bstrut \\ \hline
\Tstrut\Bstrut \textbf{Distance}  & Hausdorff Distance, q = 95th percentile\Tstrut\Bstrut & HD 95 &   \citep{huttenlocher1993comparing,kuijf2019standardized,litjens2014evaluation} &\belowbaseline[0pt]
{{\small$\begin{aligned}    
&\max\left (h_p(A,B), h_p(B,A) \right ) \textrm{ with } \\ 
& h_p(A,B) = \underset{a \in A}{P^{th}}\min\limits_{b \in B} || b - a || \\
& \textrm{ and } P = 95
\end{aligned}$}} \Tstrut\Bstrut \\
& Average Symmetric Surface Distance\Tstrut\Bstrut &ASSD  & \citep{heimann2009comparison,janssens2018fully,styner20083d} &
$\frac{\sum\limits_{x \in S} d \left (x,\hat{S} \right ) + \sum\limits_{y \in \hat{S}} d(S, y)} {|S| + |\hat{S}|} $\Tstrut\Bstrut\\
& Surface Dice at Tolerance & SDT  &   \citep{NikolovSDTheadneck,ElguindiRadiotherapy,ShusharinaSDT}&$\frac{|\hat{S} \cap B^t| + |S \cap \hat{B}^t |}{|\hat{S}| + |S|}$\Tstrut\Bstrut \\ 
& Boundary IoU & BIoU  &   \citep{cheng2021boundary} &\abovebaseline[1pt]{$\frac{|(S \cap B^d)\cap (\hat{S} \cap \hat{B}^d) |}{|(S \cap B^d)\cup (\hat{S} \cap \hat{B}^d)|}$}\Tstrut\Bstrut \\\hline
\multirow{4}{4pt}{ \textbf{Image-level \\classi\\fication$^1$}} 
& Accuracy$^2$   & ACC  &   \citep{gautam2021towards, Maier_Hein_pitfall_metric}   &   \belowbaseline[1pt]{$\frac{TP_i+TN_i}{TP_i+TN_i+FP_i+FN_i}$} \\
& $F_1$-score (equivalent to Dice)$^2$   & $F_1$-score  &   \citep{gautam2021towards,Maier_Hein_pitfall_metric}   &   \abovebaseline[0pt]{$\frac{2TP_i}{2TP_i+FP_i+FN_i}$} \\
& Sensitivity$^2$   &  Sensitivity &   \citep{Maier_Hein_pitfall_metric}    &   \abovebaseline[2pt]{$\frac{TP_i}{TP_i + FN_i}$} \\
& Specificity$^2$  & Specificity  &    \citep{Maier_Hein_pitfall_metric}    &   \abovebaseline[1pt]{$\frac{TN_i}{TN_i + FP_i}$} \\
&  Area Under the Curve$^{2}$  & AUC  &   \citep{gautam2021towards, Maier_Hein_pitfall_metric}   & $ \int\limits_{0}^{\mathrm{V_{max}}} \frac{TP_i(\mathrm{V_{threshold}})}{TP_i+TN_i+FP_i+FN_i}d \mathrm{V_{threshold}}$  \\\hline
\end{tabular}
\footnotesize{\newline $^1$ with $threshold$ $^2$subscript $i$ indicates the image-level cardinalities}
\end{table*}

A 3D image consists of a  voxel grid with width $w$, height $h$, and depth $d$. We refer to the set of voxels as $X$ with $|X| = w \times h \times d = n$.

A segmentation mask is a grid with the same shape as the image. Pixels/voxels are assigned integer values indicating the semantic class (e.g. organ, pathology) they belong to.
In the context of this publication, segmentation masks are either created manually by human experts or, with an automatic algorithm from an image.

A mask can be evaluated by the volume and location agreement of the segmented object. On a voxel level, the agreement between the reference mask, $M$, and the predicted mask, $\hat{M}$, can be measured with (i) voxel class agreement or (ii) spatial distances between corresponding voxels. 

For voxel class agreement, we use the assignment of voxels to classes, in the reference mask, as the true classes. The model’s classification for each voxel results in a predicted mask.
Let $K$ be the set of classes. We note that \hl{$K$} completely partitions the mask. That is, \hl{$M = \bigcup \limits_{k \in K} M^k$ and $\hat{M} = \bigcup \limits_{ k\in K} \hat{M}^k$.}

For a binary classification task ($k \in \{0,1\}$) a confusion matrix of four cardinalities namely $TP$, $FP$, false negatives ($FN$), and true negatives ($TN$) can be defined, where $TP + FP + TN + FN = |X|$ (Table  \ref{tab:cardinalities}).
\begin{table}[H]
\centering
\caption{\hl{Voxel-Level Cardinalities for a binary classification task in which each voxel is assigned to one of the cardinalities depending on its mapping in the reference mask, $M^1$, and predicted mask, $\hat{M}$}} \label{tab:cardinalities}
\begin{tabular}{|l|l|l|}
\hline
 & $\hat{M}^1$ & $\hat{M}^0$ \\ 
\hline
$M^1$ & $TP$    & $FN$    \\
\hline
$M^0$ & $FP$    & $TN$  \\
\hline
\end{tabular}

\end{table}
The volume, $V$, of the target object in the reference mask and volume, $\hat{V}$, in the predicted mask is defined as

\begin{align}
\label{eq:volumegt}
V &= |M^1| \times v = (TP + FN) \times v \\
\hat{V}&= |\hat{M}^1| \times v =(TP+FP)\times v
\end{align}

where $v$ refers to the physical volume of each voxel.

\label{sec:distance_based}For distance agreement, the distance between a voxel $x$ and a set of surface voxels $S^k$ is defined as

$$d(x,S^k)= min_{s^k \in S^k}d(x,s^k).$$

For a binary segmentation task with $k \in \{0,1\}$, the set of voxels, $S^1\subseteq M$, is defined as the surface voxels of the target object in the reference mask, 
and  $\hat{S}^1 \subseteq \hat{M}$ as the surface voxels of the target object in the predicted mask.

Metric definitions for common volume, overlap, and distance metrics, that were used for the experiments, can be found with their implementations on \href{https://github.com/SophieOstmeier/UncertainSmallEmpty.git}{GitHub} and in Table \ref{table:metric_definition}.

We note that the frequent class imbalance of 3D medical image segmentation ($|M^1| << |M^0|$) limits meaningful performance evaluation by any metric that includes $TN$ (Specificity, ROC, Accuracy, Kappa, etc.). Therefore, metrics that include $TN$ in their function should be avoided.
Overlap metrics measure $TP$ relative to a combination of $TP$, $FP$, and $FN$. 
Overlap metrics that exclude $TN$ are Dice, Recall, and Precision.  Volume metrics without $TN$ are VS (Volumetric Similarity) and AVD (Absolute volume Difference).

We also note that the Jaccard index $J$ (Intersection over Union, IoU) and Dice $D$ are equivalent and one can be derived from one to the other using the following formula \citep{bertels2019optimizing}.
\begin{align}
\label{eq:DiceIoU}
J(M,\hat{M})= \frac{D(M,\hat{M})}{2-D(M,\hat{M})} \\
D(M,\hat{M})= \frac{2 \times J(M,\hat{M})}{1+J(M,\hat{M})}
\end{align}
The concrete choice for either one of these metrics depends on the user or community preference\citep{Maier_Hein_pitfall_metric}.

\subsection{Surface Dice at Tolerance}
Surface Dice is an evaluation metric introduced by \citep{NikolovSDTheadneck}. It describes which portion of voxels on the surface of the target object in the predicted mask have the same spatial location as the surface voxels in the reference mask. 
For that, it classifies the surface voxels into $TP$, $FP$, and $FN$ depending on their distance to the closest surface voxel in the reference/predicted mask.
The contribution of individual voxels to these terms is weighted by the estimated surface area that it represents.
 The tolerable distance $t$ from the surface at which a voxel still counts as a TP establishes a set of border voxels  $B^t$. $t$ is a variable that needs to be set according to the clinical context and (estimated) inter-rater variability for the given segmentation task.



$$SDT = \frac{\left |\hat{S}^k \cap B^{t,k} \right| + \left | S^k \cap \hat{B}^{t,k}\right |}{\left | \hat{S}^k \right | + \left | S^k \right |}.$$

$k$ is the target object class. The tolerated distance can be set depending on the task. A possible method to choose the tolerated distance is to compute the distance between different experts as an acceptable variability (e.g. ASSD Table \ref{table:metric_definition}). This might lead to an optimization procedure with a realistic tolerance in which uncertainty within the voxel classification of the reference mask is expected and acceptable. 

The Boundary IoU with a distance $d$ proposed by \citep{cheng2021boundary} can be converted to the Surface Dice at Tolerance with tolerated distance $t$ by using Equ. \ref{eq:DiceIoU}, where $d$ and $t$ are equivalent.
\subsection{Uncertainty Score}
\label{sec:Uscore}

We develop a score to estimate the uncertainty across a set $E$ of experts across the set $C$ of cases. This score may be used as an indicator of the uncertainty in the data set. Our score is built around evaluating entropy, a measure of information contained in samples, on a target region of each image.


\hl{We index cases using $c \in C$.} We index the reference masks and membership functions by expert $e$ as $M^k_e$ and membership functions as $f^k_e (x)$. We consider the case where $k=\{0,1\}$.

Our score will require counting over experts, classes, and voxels. Let $\beta^k(c,x)$ be the function that returns the number of experts that puts voxel $x$ of case $c$ in class $k$. Formally, $\beta^k(c,x) = \left |\{e | e \in E, f^k_e (c,x) = 1\} \right |$.


We compute the U-score over the set of voxels where at least one expert classified the voxel as positive. We denote this set as $U = \left ( \bigcup\limits_{e \in E} M_e^1 \right )$.

For a case, we compute its U-score as the average, over voxels, of the expert annotation entropy of the voxel. \hl{Formally} 
\begin{align}
\textrm{U-score}= \frac{1}{|U|} \sum\limits_{x \in U} \mathrm{entropy}(x). \label{eq:Uscore}
\end{align}

With entropy computed as
\begin{align*}
    \mathrm{entropy}(c, x) &= \sum\limits_k \frac{\beta^k(c,x)}{|E|} \log \frac{\beta^k(c,x)}{|E|}.
\end{align*}

We can compute the \hl{U-score} of a dataset as the average \hl{U-score} over cases.
\subsection{Voxel-level Class Imbalances}
\hl{The class imbalance ratio ($IR$) is commonly defined as the ratio between the cardinality of the majority class and the cardinality of the minority class }\citep{zhu2020adjusting}. 

\subsubsection{Class Imbalances of Segmentation} \label{sec:classimbalancealpha}
\todo[inline]{moved from results and rephrased for better readability}
In this context of image segmentation, the $IR$ is given by $IR = \frac{|M^{majority}|}{|M^{minority}|}$.

An image segmentation task can be a perfectly balanced voxel-level binary classification problem  $|M^1| = |M^0|$. 
However, it is often the case that the target object is small relative to the image, that is  $|M^1| <<  |M^0| $ and $\frac{|M^0|}{|M^1|} >> 1$. This indicates high class imbalance. This can result in even very simple models achieving many true negatives on $|M^0|$ with a low false positive rate (Table \ref{tab:cardinalities}). 
Since the number of background voxels in medical images may vary (due to scanner settings, and image processing) we aim to control the considered background voxels in a consistent way. 
We do so by restricting the region of interest to either an organ or the immediate body cavity.
Background voxels in this region of interest are referred to as $M^{0,region}$. For example, for stroke and brain tumor this would be the brain, for the gray matter in the spinal cord this would be the total spinal cord. We then get 
\begin{align}
    IR = \frac{|M^{0,region}|}{|M^1|}.
\end{align}


\subsubsection{Image-level Class Imbalances}
\begin{figure*}[]
    \centering
    \includegraphics[width=\textwidth]{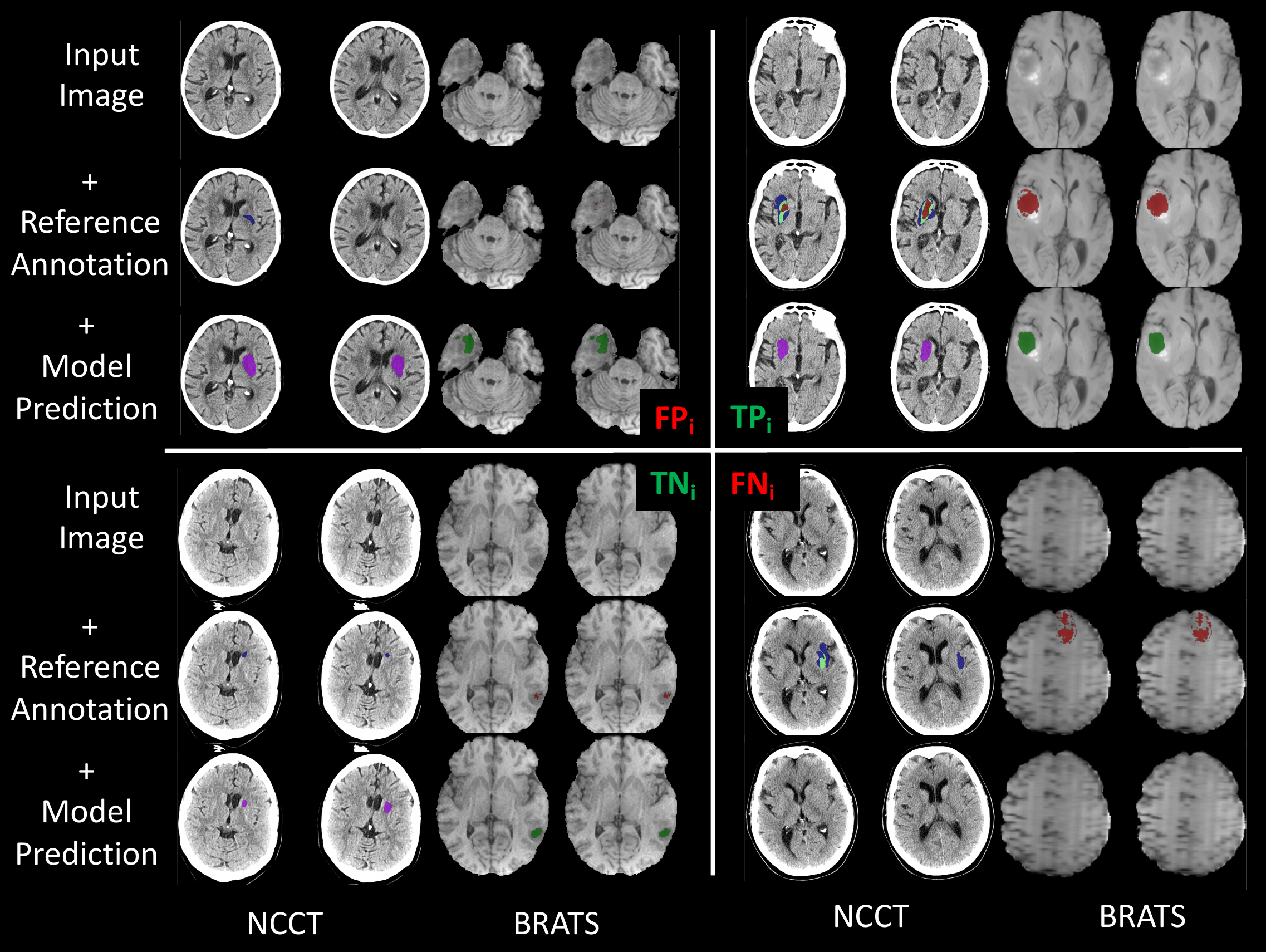}
    \caption{Example of true positives ($TP_i$), true negatives ($TN_i$), false positive ($FP_i$) and false negative ($FN_i$) cases for the NCCT and BRATS data set for a threshold of 1ml. }
    \label{fig:montage}
\end{figure*}

In the realm of image classification, we denote the class imbalance ratio as $IR_i$. When considering reference and predicted volumes that fall below a clinically reasonable $threshold$ (i.e. 1ml for the NCCT and BRATS models), the significance of segmentation performance diminishes. Images that are correctly or incorrectly classified below this $threshold$ are designated as $TN_i$ or $FN_i$ respectively. 
As a result, we derive the equation:
\begin{align}
IR_i = \frac{TP_i+FN_i}{TN_i+FP_i}
\end{align}
with an optimal value of 1. Here, the positive cases (i.e. patients with a stroke larger than 1ml), represented by $TP_i+FN_i$, serve as the majority class, while the negative cases (i.e. patients with a stroke smaller than 1ml), represented by $TN_i+FP_i$, serve as the minority class. Visual examples illustrating $TN_i, FN_i, TP_i$, and $FP_i$ can be observed in Figure \ref{fig:montage}.

\section{Methods}
\subsection{Data Sets}

\begin{table}
    \caption{Data set properties}
    \setlength\tabcolsep{1.5pt}
    \begin{NiceTabular}{|>{\bfseries}l|p[l]{1.5cm}|p[c]{1.3cm}|c|p[c]{1.3cm}|p[c]{1.3cm}|}
    \hline
        \bfseries Data set & \bfseries Target Object& \bfseries Multiple Labels & \bfseries USE$^1$ & \bfseries Positive Cases$^2$ & \bfseries Negative Cases$^3$\\\hline
        NCCT$^4$ & ischemic core & \checkmark & \checkmark &\checkmark &\checkmark\\
        BRATS 2019 & non-enhancing tumor & - & \checkmark & \checkmark & \checkmark \\
        BRATS 2019 & whole tumor & - & - & \checkmark & - \\
        Spinal Cord& gray matter & \checkmark  & - & \checkmark & - \\ 
    \hline
    \end{NiceTabular}
    \label{tab:data_sets}
    \footnotesize{\newline $^1$ USE= uncertain, small and empty reference annotations, \newline $^2$ cases with the target object present in the image, \newline $^3$ case without the target object or below a volume threshold present in the image \newline $^4$ NCCT= Non-Contrast Computed Tomography}
\end{table}
To evaluate metrics for models trained on uncertain, small, or empty reference annotations we use several data sets (Table \ref{tab:data_sets}). 

A de-identified dataset of 200 NCCT images of patients with an acute ischemic stroke from the DEFUSE3 trial \citep{Albers6to16} was provided to three neuroradiologists 4, 4, and 5 years of experience in neuroradiology (B.V.,A.M.,J.J.H.) (study design \url{https://clinicaltrials.gov/ct2/show/NCT02586415}). The experts were instructed to segment abnormal hypodensity on the NCCT that corresponds to acute ischemic brain injury.
Detailed instructions and videos, as well as an oral explanation of the task, were given. 
Any missed lesions or missed slices were not corrected. The experts' masks were fused by a majority vote to form the reference mask.
In addition, 156 institutional NCCT images were added of patients who were scanned with suspicion of stroke but were confirmed on follow-up Diffusion-weighted MR imaging not to have a stroke.

The BRATS 2019 public data set was used to reproduce and compare results and included 345 MRIs of high and low-grade glioma patients \citep{BakasBRATS19,BakasBRATS2019_2,MenzeBRATS19_3}. One to four experts segmented the brain tumors followed by a consensus procedure. The reference masks had four target objects; "background", "edema", "non-enhancing", and "enhancing". We used this data set to train two segmentation tasks (i) with the target object "non-enhancing" tumor on only T1 and (ii) with the target object whole tumor, defined as the union over "edema", "non-enhancing", and "enhancing" target objects, on T1, T1 contrast-enhanced, T2-Flair and T2.

The Spinal cord data set is a public data set with 40 \hl{annotated} MRIs of 40 healthy patients from 4 different hospitals and annotated by 4 experts per case. The annotations include the white and gray matter of the spinal cord on T2\citep{prados2017spinal}. The experts' masks
were fused by a majority vote to form the reference mask.

\subsection{Data Partition}
\begin{figure}
    \centering
    \includegraphics{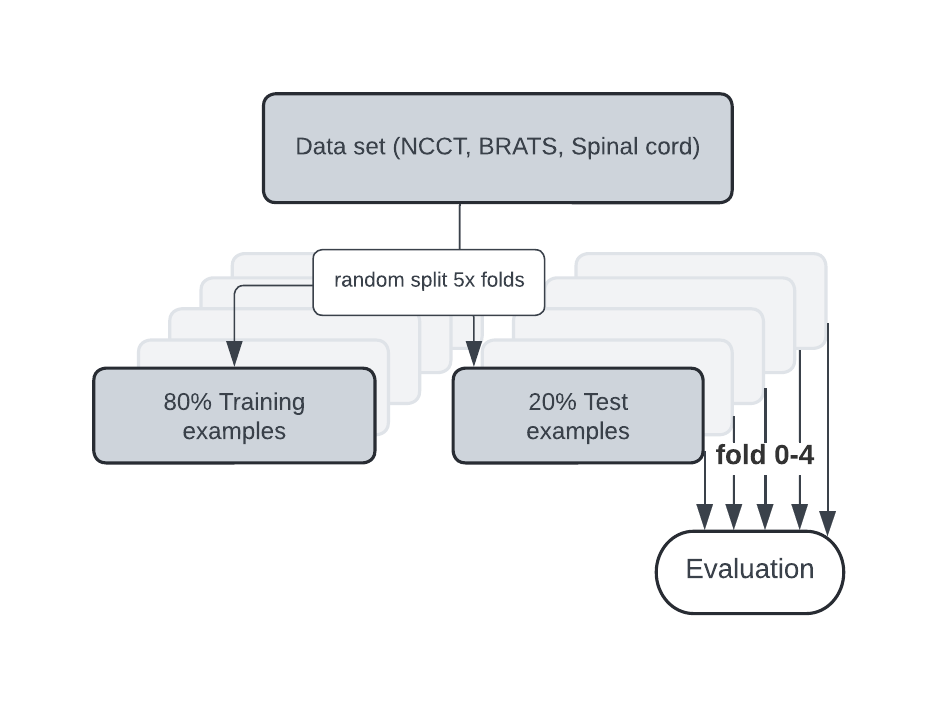}
    \caption{Data Sampling and Partition of 5-fold-Cross-Validation}
    \label{fig:split}
\end{figure}

For each segmentation tasks the cases were randomly divided into 5 folds that consisted of 80\% training and 20\% test examples. The default self-configured nnUNet was used to train all folds for each segmentation task.
All analyses were done on the aggregated 5 test sets for each segmentation task (Supplemental material, Figure \ref{fig:split}).

All models shared the same training schedule with 500 epochs, Stochastic gradient descent with Nestov momentum of 0.99, the initial learning rate of 0.01 with linear decay, and oversampling of 33\% for the target lesion.

\subsection{Models}
\subsubsection{Deep Learning models}
We chose the 3D full-resolution nnUNet as our deep learning framework \citep{IsenseennUNet}. For fairness and ease of comparability, we let all models undergo the same training schedule and did not modify hyperparameters.

The default configured model included a patch size of ($1\times 28 \times 512 \times 512$) and spacing of (3.00, 0.45, 0.45), Dice and Cross-Entropy loss function, seven stages, two 3D convolutions per stage and a leaky ReLU as activation function. 

For the NCCT ischemic core segmentation task, the model input was the NCCT image (356 cases) to output a predicted mask for ischemic brain tissue.

For the first BRATS 2019 segmentation task, the model input encompasses only the T1 image (345 cases) to simulate a lower signal-to-noise ratio and the output was the predicted mask for the non-enhancing tumor. 
For the second BRATS 2019 segmentation task, the model input included all available MR sequences images (345 cases) to output the predicted mask of all tumor parts. 

For the Spinal cord gray matter segmentation task (40 cases), the model input was a T2 image to output a predicted mask of gray matter.

\subsubsection{Random model}
\label{sec:random_model}
\todo[inline]{moved here from results and rephrased for better readability}
To demonstrate that this trend is independent of the model, we analyze the behavior of the Dice on a model that randomly labels voxels.

Our objective is to investigate the impact of target object size on the Dice score.
As the class imbalance ratio increases, the Dice score tends to decrease, creating difficulties in comparing model performance across different levels of class imbalance. 
However, we aim to demonstrate that this relationship is also a general property not tied to a single task. 

Consider a binary segmentation model that decides each voxel's membership in the predicted mask randomly with a biased coin toss \label{def:randommodel}.
We will show this by deriving the expected Dice score $E_D$ for images drawn from a random model and showing that the trend observed empirically matches the trend in this theoretical model (Supplemental material \ref{app:randomdice}.).
It is cleaner to parameterize this random model using the expected portion of voxels that are positive. We refer to this as $p$ and note that it can be directly computed from the class imbalance ratio $p = \frac{1}{1 + IR}$. 

\subsection{Evaluation Tool}
\label{sec:evaltool}
All evaluations were performed using the \href{https://github.com/SophieOstmeier/UncertainSmallEmpty.git}{USE-Evaluator} inspired by \citep{NikolovSDTheadneck,IsenseennUNet} (Table \ref{table:metric_definition}). 
The source code can be applied to folders with reference annotation and prediction mask in .nii.gz format and produces a .xslx file with sheets for all studies, the means, medians, and image-level classification with bootstrapped 95\% confidence interval. 
A threshold flag can be set as a lower volume threshold for the segmentation and image-level classification evaluation. 
If the reference or predicted volume is below the threshold, a case is excluded from the segmentation evaluation but included as a negative case for the image-level classification evaluation.

\subsection{Evaluation of Reference Annotations}
We analyzed the variability among different experts' annotations masks, available for the NCCT and the Spinal cord data set, with the evaluation tool described in Section \ref{sec:evaltool}. 
To estimate uncertainty we compute the U-score (Equ. \ref{eq:Uscore}) \hl{and the median inter-expert agreement, and the median agreement to the majority vote (majority-expert) with the metrics presented in Table} \ref{table:metric_definition}.

\subsection{Evaluation of Model Performance}
Performances were measured with the evaluation tool (Section \ref{sec:evaltool}) with a threshold of 1ml for the NCCT and BRATS 2019 data sets. For other medical applications, this might depend on the clinical task the model is trained on.
With the evaluator tool, this threshold can be easily changed. 
For the Spinal cord data set, we did not set a threshold, because the clinical concern in healthy populations would not be about the non-existence vs. existence of gray matter in the spinal cord. 

\subsection{Evaluation of Metrics}

We evaluate the segmentation metrics by correlation to uncertainty among the expert's masks, independence from reference volume, the reward of volumetric and location expert-model agreement, and evaluation of correct classification of empty reference masks or small reference volumes cases using the R package corrplot (Version 0.92).

To compute $IR$ and $p$ for the stylized model, we defined the $region$s as the entire brain for the NCCT and BRATS datasets, and the entire spinal cord for the Spinal Cord dataset (Section \ref{sec:classimbalancealpha}). The BET\_CT was used to extract the brain on NCCT according to  \citep{SchellMRT_BET}. \hl{For the extraction of the brain on MRI, the HD\_BET was applied }\citep{isensee2019automated}.\hl{ For extraction of the spinal cord, the union of the gray and white matter in the majority vote reference mask was used.}

For the evaluation of empty reference and predicted masks, we explore possible image-classification metrics and \hl{their relationship to $IR_i$, where we refer to $p_i = \frac{1}{1 + IR_i}$.}

\section{Results and Discussion} 
\todo[inline]{rephrased paragraph, fused results, and discussion to offer a link between theory.}
\hl{In this section, we will examine the relationship between metric values and varying prevalence of uncertain, small, or empty reference annotations.\\
In Section }\ref{sec:eval_reference_annotations} \hl{we measure uncertainty in reference annotations. We conduct empirical validation of the U-score across data sets and its correlation with inter-expert variability and consensus among the majority of experts. \\
In Section}\ref{sec:eval_performance} \hl{we analyze all models' performances with each metric across data sets in order to provide a first indication of trends between dataset properties and metric values that we explore in further detail in the following section. \\
In Section}\ref{sec:eval_metrics} \hl{we use the correlation of metric values to provide empirical evidence of the link between the uncertain, small, and empty reference annotations and the metric values. \\
For the Dice metric, we demonstrate that the link is even more general by illustrating that the relationship found empirically is present in the evaluation of a stylized theoretical model (Section }\ref{sec:eval_dice}).\\
\hl{Finally, we explore trends in image-classification metrics in section }\ref{sec:image_classification_guidelines}.

Upon negative tests for normal distribution, results for each metric are shown as medians with 95\% confidence interval (bootstrapped, 1000 repetitions), and the correlations are reported as Spearman's rank correlation coefficient.

\subsection{Evaluation of Reference Annotations of Experts}
\label{sec:eval_reference_annotations}
Variability in reference annotations can impact the model's segmentation performance and solutions have been discussed \citep{karimi2020deep}. 
However, we focus on a better choice of evaluation techniques to enhance the clinical applicability of segmentation models.
In this regard, we first propose the introduction of the \hl{U-score} as a measure of uncertainty for reference annotations (Section \ref{sec:uncertainty}). We found an overall median U-score is significantly different between the NCCT ischemic core and the Spinal cord gray matter segmentation task (0.87 ± 0.05 vs. 0.39 ± 0.02, respectively). These findings are consistent with common measures such as inter-expert and majority-expert agreement (supplemental material, Table \ref{tab:interexpert})\citep{10054393}.
Inter-expert and majority-expert agreements use pairwise expert comparison and rely on common segmentation metrics to indirectly estimate uncertainty in reference annotations. The U-score directly measures uncertainty.

We found varying distributions of reference volumes across the studied data sets (median (IQR) volume 6 (2-21)ml, 10 (4-25)ml, 89 (48-146)ml and 0.7 (0.3-1.1)ml for NCCT, BRATS 2019 non-enhancing tumor part segmentation task, BRATS 2019 whole tumor segmentation task and Spinal Cord gray matter segmentation, respectively. 

We further conduct correlation analyses between the U-score and reference volumes to common metrics outlined in Section \ref{sec:eval_metrics}.

\subsection{Evaluation of Segmentation and Image Classification Performance}
\label{sec:eval_performance}

\begin{table*}[t!]
\caption{Results of Segmentation Task Performance $^1$}
\begin{tabular}{|>{\bfseries}l|>{\bfseries}l|rl|rl|rl|rl|}
\hline

\textbf{Categories}&\textbf{Metrics$^2$}& \multicolumn{2}{l|}{\textbf{NCCT}} & \multicolumn{2}{l|}{\textbf{BRATS 2019}}&\multicolumn{2}{l|}{\textbf{BRATS 2019}}&
\multicolumn{2}{l|}{\textbf{Spinal Cord}}\\

&& \multicolumn{2}{l|}{\textbf{Ischemic core}} & \multicolumn{2}{l|}{\textbf{Non-enhancing}}&\multicolumn{2}{l|}{\textbf{Whole tumor}}& \multicolumn{2}{l|}{\textbf{Gray matter}}  \\

&& \multicolumn{2}{l|}{\textbf{}} & \multicolumn{2}{l|}{\textbf{tumor}}&\multicolumn{2}{l|}{\textbf{}}& \multicolumn{2}{l|}{\textbf{}}  \\

  \hline
Volume & VS & 0.58 & ± 0.09 & 0.78 & ± 0.03 & 0.97 & ± 0 & 0.92 & ± 0.02 \\ 
   & AVD & 4.48 & ± 1.17 & 4.41 & ± 0.95 & 4.95 & ± 0.85 & 0.08 & ± 0.04 \\ \hline
  Overlap & Dice & 0.56 & ± 0.04 & 0.60 & ± 0.04 & 0.93 & ± 0.01 & 0.83 & ± 0.02 \\ 
   & Precision & 0.69 & ± 0.11 & 0.66 & ± 0.06 & 0.94 & ± 0.01 & 0.89 & ± 0.03 \\ 
   & Recall & 0.20 & ± 0.08 & 0.58 & ± 0.05 & 0.93 & ± 0.01 & 0.79 & ± 0.03 \\ \hline
  Distance &ASSD & 2.34 & ± 0.41 & 2.50 & ± 0.15 & 0.94 & ± 0.07 & 0.12 & ± 0.02 \\ 
   & HD 95 & 8.30 & ± 1.51 & 8.00 & ± 0.55 & 2.87 & ± 0.3 & 0.50 & ± 0.05 \\ 
   & SDT small $^3$& 0.61 & ± 0.05 & 0.56 & ± 0.03 & 0.93 & ± 0.01 & 0.84 & ± 0.08 \\ 
   & SDT large $^4$& 0.86 & ± 0.03 & 0.85 & ± 0.02 & 0.99 & ± 0 & 0.84 & ± 0.08 \\ 
   \hline
\end{tabular}
\footnotesize{\newline $^1$ median ± 95\% Confidence Interval (bootstrapped)\newline$^2$ VS = Volumetric Similarity, AVD = Absolute Volume Difference,ASSD} = Average Surface Distance, HD 95 = Hausdorff Distance 95th percentile, SDT = Surface Dice at Tolerance \newline $^3$ Surface Dice at Tolerance with 2mm for NCCT and BRATS 2019 models and 0.05mm for the Spinal Cord model \newline $^4$ Surface Dice at Tolerance with 5mm for NCCT and BRATS 2019 models and 0.1mm for the Spinal Cord model\label{tab:segmentation_results}
\end{table*}

The performance of the NCCT ischemic core and BRATS 2019 non-enhancing tumor models, trained using uncertain, small, and empty reference annotations, shows similar results across volume, overlap, and distance metrics. However, the BRATS 2019 whole tumor and Spinal cord model, trained on larger and more certain reference annotations, consistently outperforms both the NCCT and BRATS 2019 non-enhancing tumor model (Table \ref{tab:segmentation_results}).

\begin{figure}
    \centering
    \includegraphics[width=0.48\textwidth]{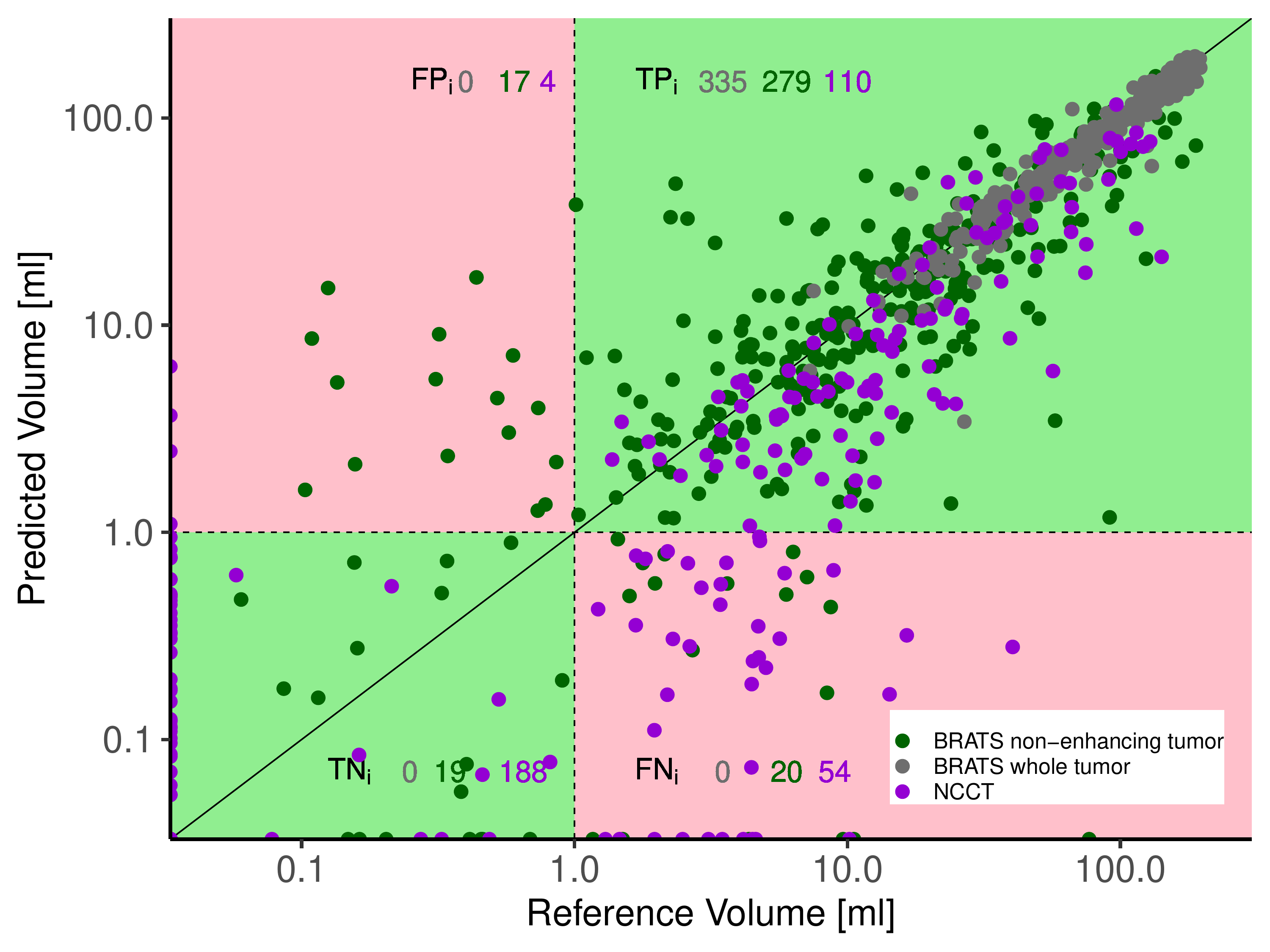}
    \caption{Scatter plot with log-scale and confusion matrix with a volume threshold of 1ml dividing $TP$ and $TN$ from $FP$ and $FN$. For the NCCT data set(violet points), almost all incorrectly classified cases are too small, namely $FN$, whereas for the BRATS non-enhancing tumor data set the opposite is the case. None of the cases of BRATS whole tumor are incorrectly classified. \label{fig:scatter}}
\end{figure}

For image classification, the total number of cases with reference volumes $<1$ml is 192 cases for the NCCT ischemic core segmentation task, 36 cases for the BRATS 2019 non-enhancing tumor part segmentation task and 0 cases for the BRATS 2019 whole tumor \label{sec:empty_cases}. 
We visualize these class distributions and report the confusion matrix (Figure \ref{fig:scatter})\citep{Maier_Hein_pitfall_metric}. Reference Volumes $>$1ml cluster around the identity line, whereas references $<$1ml are more spread.
Sensitivity, F1-score and ACC of the NCCT model are lower compared to the BRATS non-enhancing tumor models, however, the AUC and Specificity are higher for the NCCT models (Table \ref{tab:classification_results}).
Further analysis between data set properties and common image classification metrics analysis are summarized in Section \ref{sec:image_classification_guidelines}.

\begin{table*}
\caption{Results of Image-Classification Task$^1$}
\begin{tabular}{|>{\bfseries}l|>{\bfseries}l|rl|rl|rl|rl|}
\hline
\textbf{Categories}&\textbf{Metrics$^2$}& \multicolumn{2}{l|}{\textbf{NCCT}} & \multicolumn{2}{l|}{\textbf{BRATS 2019}}&\multicolumn{2}{l|}{\textbf{BRATS 2019}}&
\multicolumn{2}{l|}{\textbf{Spinal Cord}}\\

&& \multicolumn{2}{l|}{\textbf{Ischemic core}} & \multicolumn{2}{l|}{\textbf{Non-enhancing}}&\multicolumn{2}{l|}{\textbf{Whole tumor}}& \multicolumn{2}{l|}{\textbf{\hl{Gray matter$^4$}}}  \\

&& \multicolumn{2}{l|}{\textbf{}} & \multicolumn{2}{l|}{\textbf{tumor}}&\multicolumn{2}{l|}{\textbf{}}& \multicolumn{2}{l|}{\textbf{}}  \\
  \hline
Class imbalance & $p_i$ & 0.46 && 0.89 && 1.00 && \hl{-} & \\ \hline
Image-level & Sensitivity & 0.67 & ± 0.04 & 0.93 & ± 0.01 & 1.00 & ± 0 & \hl{-} &  \\ 
    & Specificity & 0.98 & ± 0.01 & 0.53 & ± 0.09 &  &   & \hl{-}  &  \\ 
    & F$_1$-score & 0.79 & ± 0.03 & 0.94 & ± 0.01 & 1.00 & ± 0 & \hl{-} & \\ 
    & ACC & 0.84 & ± 0.02 & 0.89 & ± 0.02 & 1.00 & ± 0 & \hl{-} &  \\ 
    & AUC & 0.91 & ± 0.02 & 0.86 & ± 0.03 &  &  &  \hl{-}  &  \\ 
   \hline
\end{tabular}
\footnotesize{\newline $^1$ median ± 95\% Confidence Interval (bootstrapped)\newline $^2$ 1ml threshold, \newline$^3$ ACC=Accuracy, AUC=Area under the Curve \newline $^4$healthy cohort, no threshold for pathology set
\label{tab:classification_results}}
\end{table*}
\subsection{Evaluation of Segmentation Metrics}
\label{sec:eval_metrics} 

\begin{figure}[t]
\centering
\includegraphics[width=0.48\textwidth]{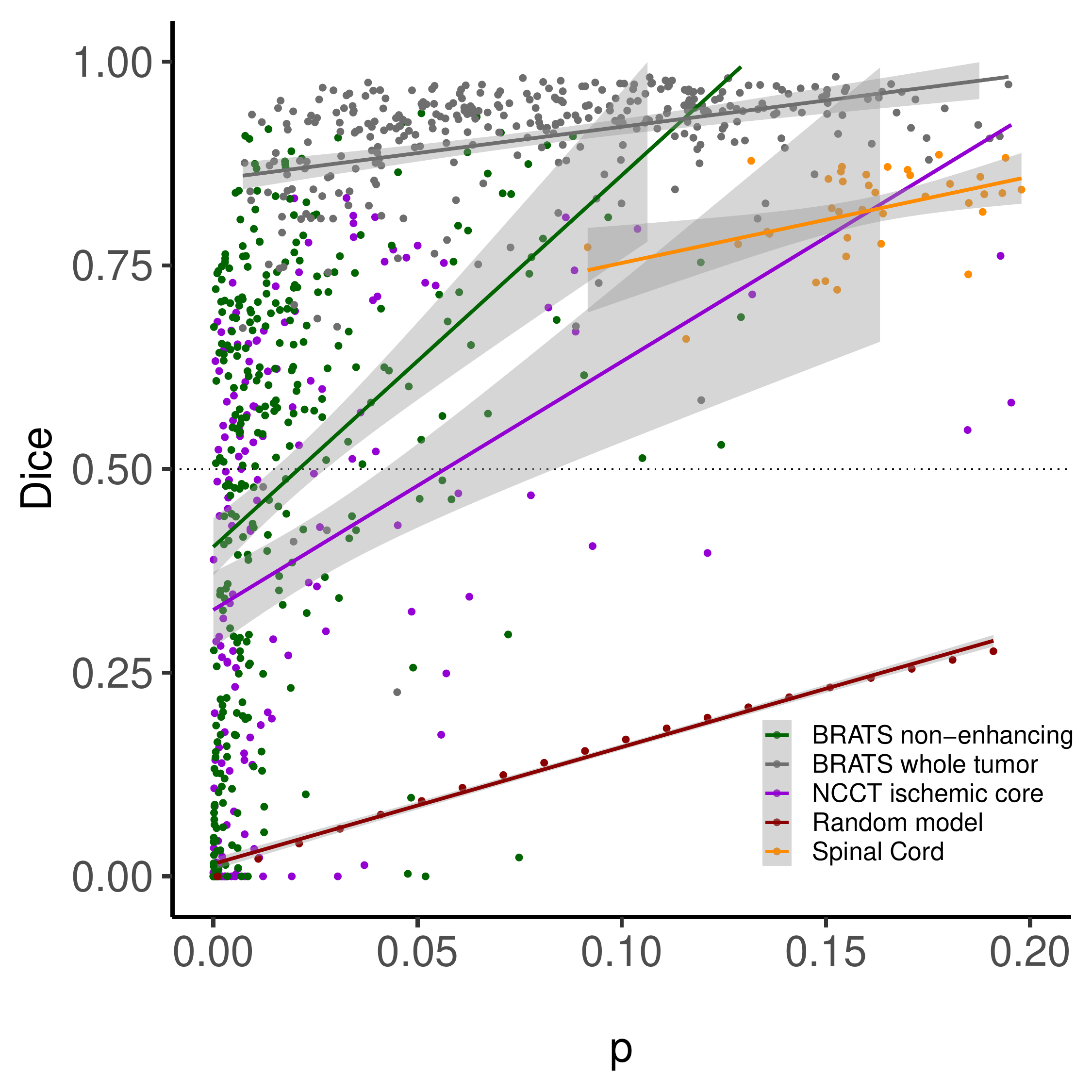}
\caption{
Dot plot with regression lines for the Dice over class imbalance $p$ for all segmentation models, where $p = \frac{1}{1+IR}$. The gray areas represent 95\% confidence intervals. The dark red dots and line represent the random model with the expected Dice $E_D$ defined \hyperref[def:randommodel]{here}.  The dashed line indicates the expected Dice $E_D$ for a balanced reference mask. \label{fig:alpha}}
\vspace{-1\baselineskip}
\end{figure}

\begin{figure*}
    \centering
    \begin{subfigure}[b]{9.0cm}
    \centering
    \includegraphics[width=\textwidth]{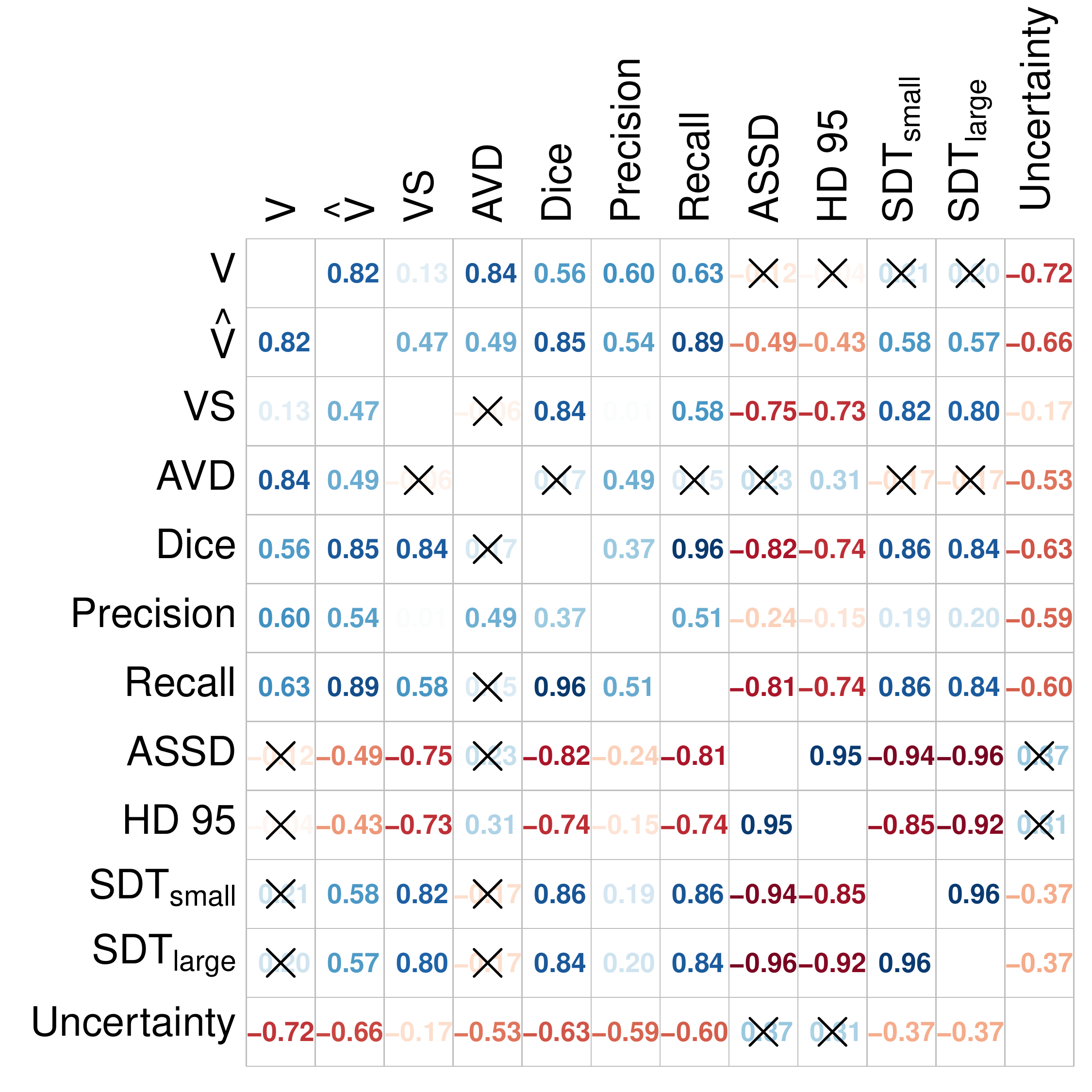}
    \caption{\hl{Uncertain, small and empty reference annotations (NCCT stroke): The overlap metrics (Dice, Recall and Precision) and AVD have a strong negative correlation to Uncertainty and positive correlation to Reference Volume, Distance metrics (ASSD, HD 95, SDT) show insignificant correlation, respectively.}}
    \end{subfigure}
    \hspace{0.5em}
    \begin{subfigure}[b]{9.0cm}
    \centering
    \includegraphics[width=\textwidth]{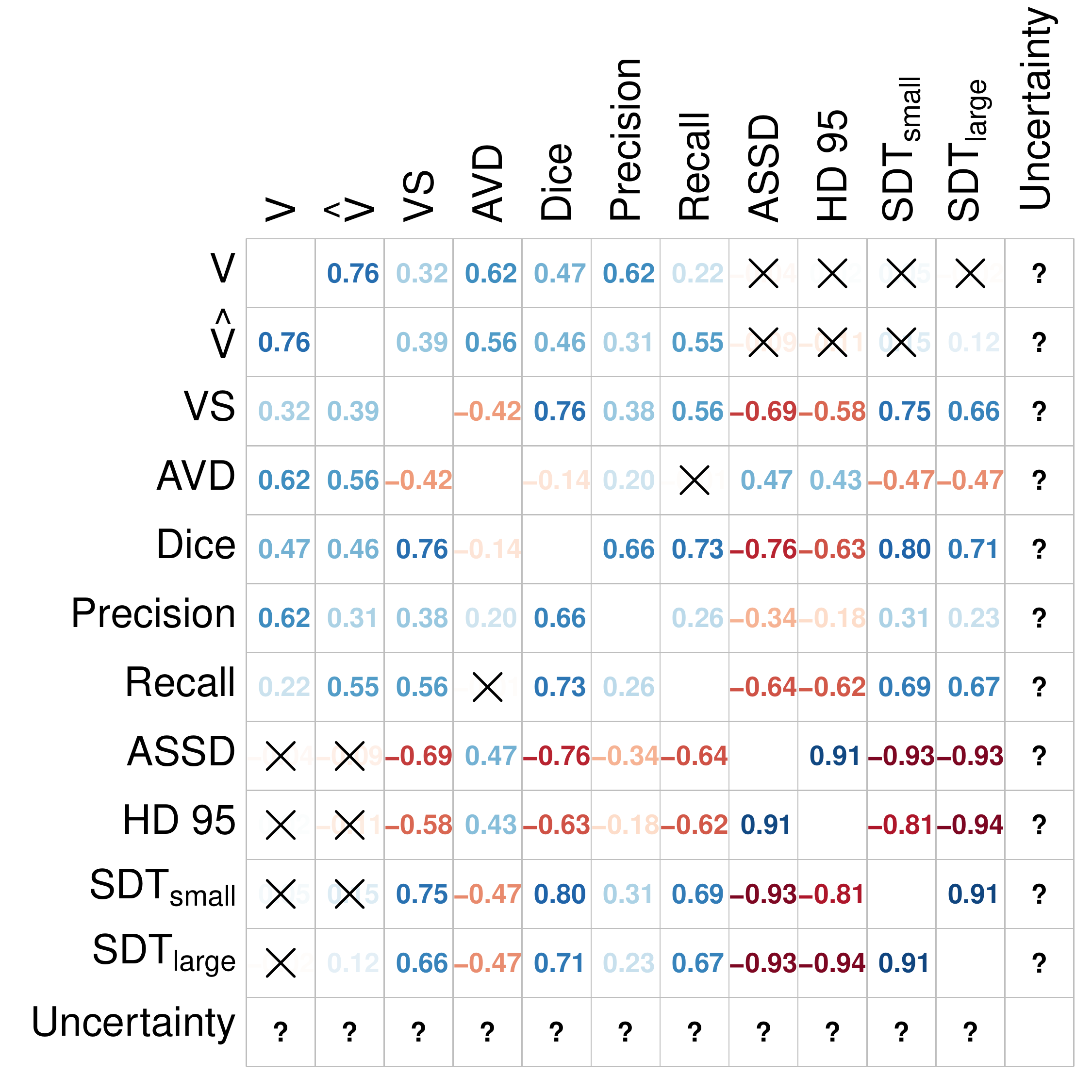}
    \hspace{1\baselineskip}
    \caption{\hl{Uncertain, small and empty reference annotations (BRATS 2019 non-enhancing tumor): Like in (a), overlap metrics (Dice, Recall, and Precision) and AVD demonstrate a strong positive correlation with the reference volume (V), while distance metrics (ASSD, HD 95, SDT) exhibit insignificant correlation.}}
    \end{subfigure}
    \begin{subfigure}[b]{9.0cm}
    \centering
    \includegraphics[width=\textwidth]{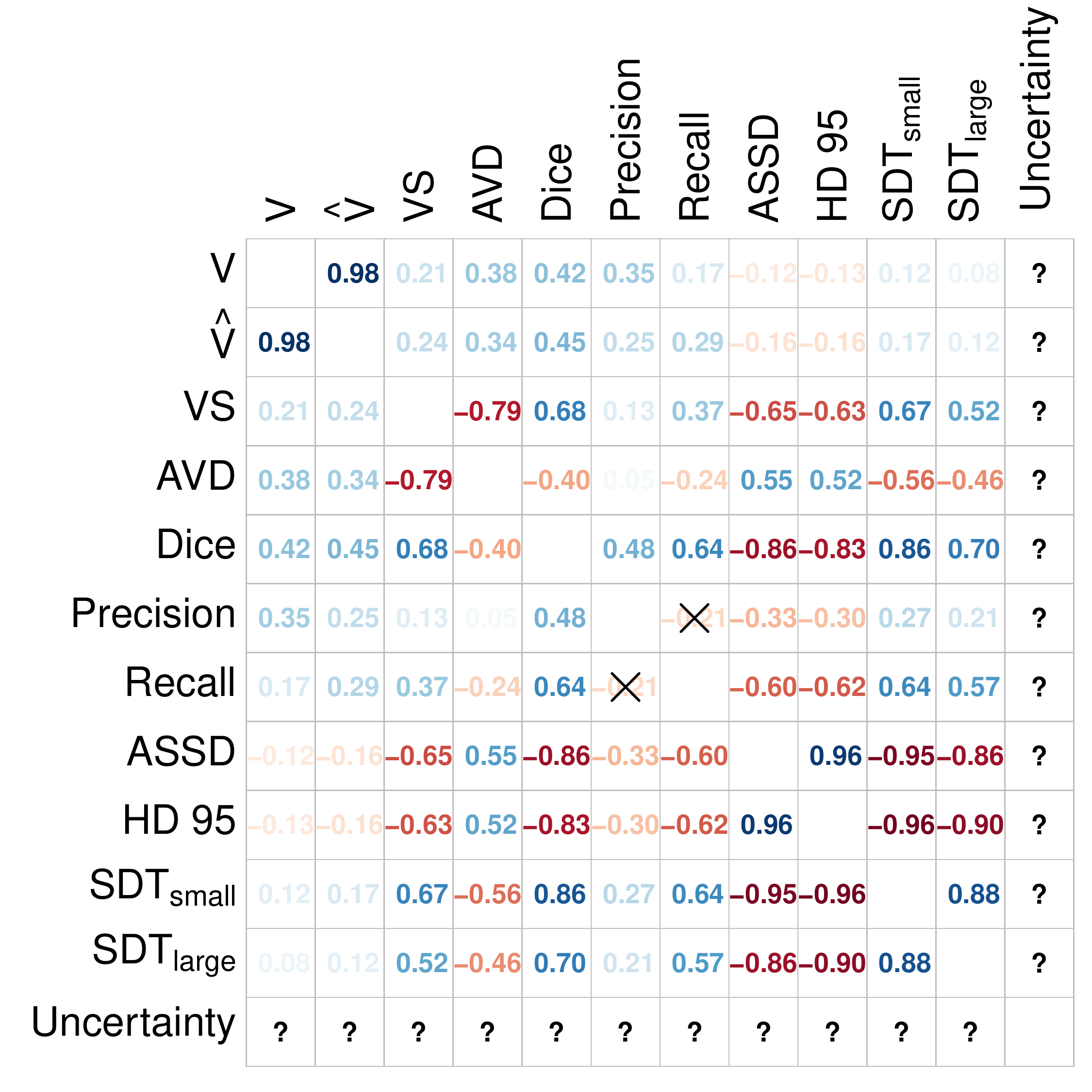}
    \caption{\hl{Certain and Large reference annotations (BRATS 2019 whole tumor): Low correlation between the volumes to metric values.}
    \\
    \\}
    \end{subfigure}
    \hspace{0.5em}
    \begin{subfigure}[b]{9.0cm}
    \centering
    \includegraphics[width=\textwidth]{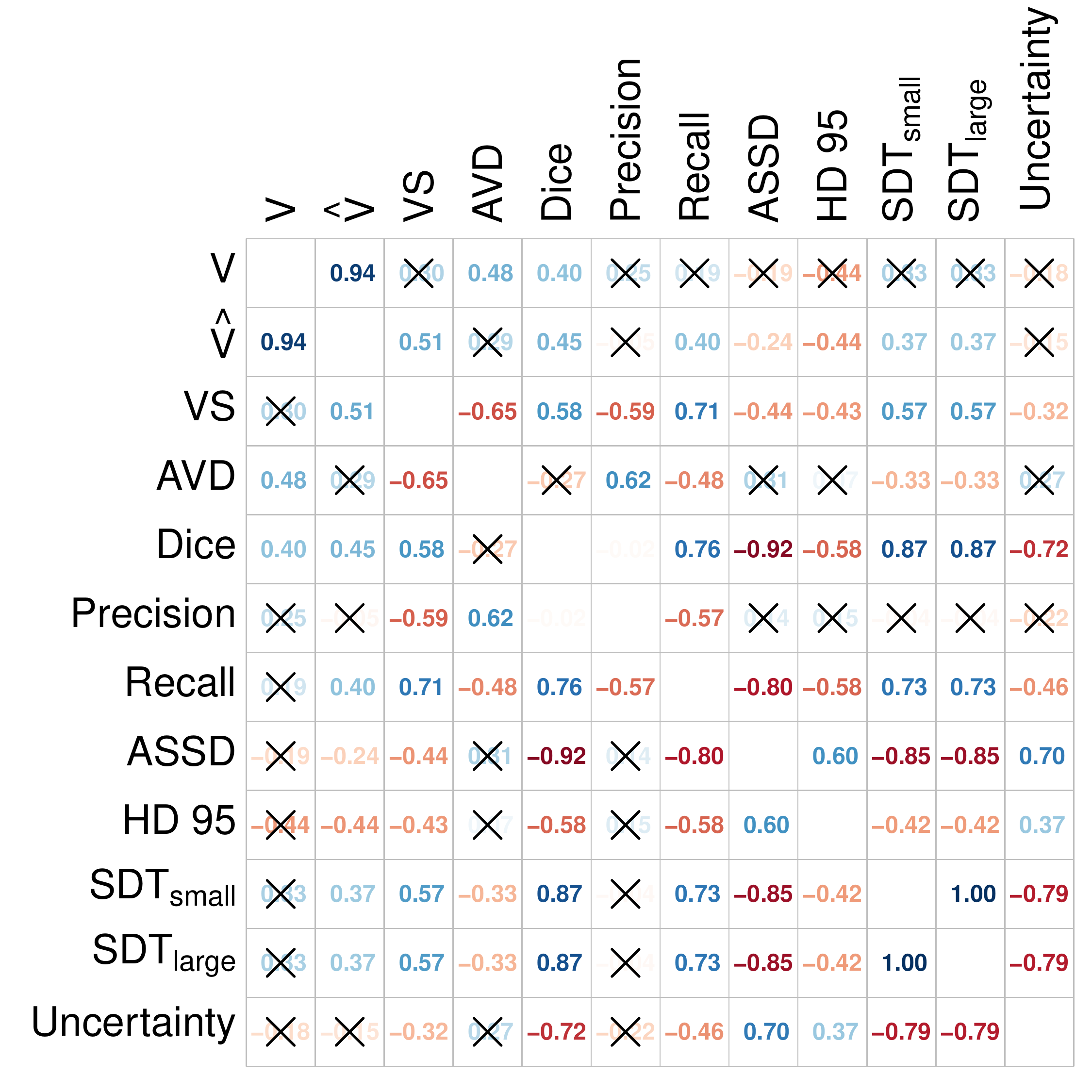}
    \caption{\hl{Certain and small reference annotations (Spinal Cord gray matter): 
With the exception of Dice and AVD, there is a lack of correlation between the metrics and the reference volumes. Dice and SDT exhibit a high correlation with uncertainty.}}
    \end{subfigure}
    \centering
    \caption{\hl{Correlation matrices of Spearman coefficient for data sets and metrics. X indicates insignificant correlations with $p>0.05$. Overall correlation patterns among metrics (e.g. Dice and SDT) remain similar over the data sets. The correlation between Dice and uncertainty, as well as the reference volume, is reproducible in all datasets, albeit to varying degrees.}}
    \label{fig:correlation_matrix_numbers_1}
\end{figure*}

\hl{We use the relationship between metrics and dataset properties to identify evaluation strategies robust of the presence to uncertain, small or empty reference annotations. 
These recommendations are backed by the empirical data} (Figure\ref{fig:correlation_matrix_numbers_1}) \hl{and we provide an intuition of how the given formula provides the observed effect} (Table\ref{table:metric_definition}).

\hl{We categorize the segmentation metrics according to volume, overlap, or distance agreement. We then analyze every segmentation metric based on the characteristics outlined in section} \ref{sec:clinical_value}. \hl{If not otherwise specified, all numbers in this section refer to the Spearman correlation coefficients presented in Figure }\ref{fig:correlation_matrix_numbers_1}. \hl{A summary of the core results and guidelines for the choice of metrics are provided in Table }\ref{tab:guidelines}.

\subsubsection{Volume Agreement}
\textbf{VS:}

\emph{Robustness toward Uncertainty and Independence from Reference Volume:}

Conceptually, VS allows location variability of reference volumes (Table \ref{tab:guidelines}), because $FN$ and $FP$ voxels can be anywhere in the image without an influence on the value of VS. 
This characteristic becomes particularly valuable when dealing with uncertain reference annotations.
Assuming that a source of $FP$ and $FN$ is uncertainty (see Section \ref{sec:uncertainty}); VS does not penalize uncertainty as long as their difference has a linear relationship to reference volumes. This is because VS normalizes to the sum of the reference and predicted volume. 
Our findings support this with a low correlation to uncertainty (-0.17 and -0.32 for NCCT and Spinal Cord) and reference volume (across all data set below 0.25). 
We conclude that VS value is less driven by uncertainty or reference volume.

\emph{Reward of Volume and Location Agreement:}

VS does not reward location agreement since VS measures the relative relationship between $FP$ and $FN$ rather than their distance. VS is therefore suitable if volume agreement is the major clinical concern, as for some applications in neuroimaging, like stroke \citep{PowersManagement}. 

In theory, VS may be less appropriate for clinical datasets and segmentation tasks that heavily rely on spatial information, such as those involving multiple sclerosis \citep{filippi2019assessment}. However, our observations indicate a consistent moderate to strong correlation with overlap metrics (e.g., Dice coefficient ranging from 0.58 to 0.84) and distance metrics (e.g., SDT$_{small}$ ranging from 0.57 to 0.82), particularly in datasets where reference annotations are uncertain and small in size.

\emph{Reward of Agreement of Emptiness:}

For cases with empty references and predicted masks, VS returns the optimal value of 1. 
Therefore, VS is suitable for data sets with expected empty reference masks. 
Nevertheless, we recommend setting a threshold for very small volumes ($<$1ml), because the frequency of empty reference or predicted masks could screw the distribution of values compared to other metrics.

\textbf{AVD:}

Our findings suggest, no advantage of AVD over VS for data sets with a small median of reference volumes and uncertainty. 

In contrast to VS, AVD does not normalize to the sum of reference and predicted volumes. Larger reference volumes have potentially larger volume differences, resulting in a notably positive correlation between AVD and reference volumes across all data sets. 
In datasets with a wide spread of reference volumes (Figure \ref{fig:distribution}), it is unclear whether a reduction of AVD as a metric leads to slightly improved performance for large reference volumes or substantially for small reference volumes. 
This ambiguity can introduce bias when comparing model performance within and across datasets, as evidenced by inconsistent correlation patterns with overlap and distance agreement metrics in our correlation analysis.

\subsubsection{Overlap Agreement }
\textbf{Dice:}
\label{sec:eval_dice}

\emph{Robustness toward Uncertainty:}\\
We observed that the Dice correlates more with uncertainty compared to other metrics (-0.62 to -0.72). This indicates that the Dice value is influenced not only by the extent of overlap but also by the level of uncertainty.

In a theoretical context, let's consider two scenarios. 
In the best-case scenario, a model outperforms the experts (as determined by the majority vote of reference annotation) by correctly classifying voxels. In this ideal situation, all $FP$ are $TP$, and all $FN$ are $TN$. 
However, the denominator contains the sum of $|M^1|=TP+FN$ and $|\hat{M}^1|=TP+FP$ (Table \ref{table:metric_definition}) and would disproportionately increase and lead to a lower Dice value. As a result, the performance of the models is underestimated.
In the worst-case scenario, a model is inferior to the experts in classifying voxels correctly; all $FP$ are truly $FP$ and all $FN$ are truly $FN$. The Dice value does not change. As a result, Dice is biased toward the worst-case scenario. 
Hence, the Dice over-penalizes overlap disagreement in the presence of uncertainty between the experts' masks with a lower value.

\emph{Independence from Reference Volume:}

In our study, we consistently observed a positive correlation between the reference volume and the $IR$ across all datasets, ranging from 0.40 to 0.56, with the NCCT dataset exhibiting the highest correlation. 
We hypothesized that the size of the target object affects the Dice value. More specifically, we investigated how the $IR$ impacts the likelihood of a voxel being classified as $TP$ because the Dice primarily rewards accurate voxel assignment to $TP$ ($\frac{2TP}{2TP+FP+FN}$).
To validate our hypothesis, we analyzed the Dice value on a random model using the parameter $p$ (Section \ref{sec:classimbalancealpha}). 
The value of $p$ can be directly calculated from the $IR$ and represents the probability of a voxel in the prediction mask being classified as belonging to the target object class (see Section \ref{sec:random_model}). 
We plot the Dice curve of the random model (dark red line) and compared it to all data sets (Figure \ref{fig:alpha}). 
If $p$ is very low at 0.01 (1\% of the brain), then the expected Dice of the random model is 0.02. If $p$ is 0.5 (50\% of the brain), the expected Dice is much higher at 0.5 (dashed line). 
The regression lines for the random model, NCCT, and BRATS non-enhancing tumor show a positive monotonic tendency of the Dice values with higher $p$. This behavior is also present in BRATS 2019 whole tumor and Spinal cord models with larger $p$, but less (shallower slope of gray and orange lines).

We infer that a high imbalance ratio is more likely to produce lower Dice values. 
Location and volume errors for small reference annotations may be more penalized than larger reference annotations, making the Dice a sub-optimal choice of metric for data sets with small reference annotations and a wide distribution of reference volumes. 

\emph{Reward of Volume and Location Agreement:}\\
The numerator of the Dice, which comprises $2TP$, represents the voxels assigned to both the reference and prediction masks. The maximization of this value occurs when there is a high agreement in terms of both location and volume between the masks. 
We empirically see that the Dice rewards of volume and location agreement with a consistent, moderate to strong correlation to VS and distance metrics across all data sets.

\emph{Reward of Agreement of Emptiness:}\\
The Dice does not reward the agreement of emptiness between the reference and predicted mask, but returns "NaN". We found a high number of cases in the NCCT data set with a Dice value of 0. 
Investigation showed, that the Dice is zero if target objects are right next to each other and also zero if they are far from each other, especially for small reference volumes. 
This may lead to a disproportionate count of cases with Dice equal to zero.
Depending on the clinical context, very small reference volumes (i.e. $<$1 ml) may be excluded from the evaluation of segmentation metrics. This is done to avoid introducing bias to the overall performance without obtaining meaningful information.

Instead, we suggest image-classification metrics to evaluate very small reference volumes or empty reference annotations masks.
For example, a case with $V<1$ml may be better evaluated by image classification metrics than by a segmentation metric.
We implemented this idea with the USE-Evaluator, where a lower volume threshold can be set that will exclude studies with $V< threshold$ and automatically initializes an image-classification evaluation.

\textbf{Recall and Precision:}

Overall, Recall and Precision show similar behavior compared to the Dice, but only capture certain aspects of overlap agreement, and should be evaluated with other segmentation metrics and in the context of the clinical question. 

They differ in their consideration of $FP$ and $FN$ in the denominator. Precision rewards $TP$ relative to the predicted volume, $TP+FP = |\hat{M}^1|$, and Recall $TP$ relative to the reference volume, $TP+FN = |M^1|$ (Table \ref{table:metric_definition}).

Especially Recall showed a correlation to uncertainty in the NCCT and Spinal cord data set (-0.60 and -0.43).
One can argue that in the setting of high-class imbalance, the models learn to classify voxels with high entropy less frequently to $|\hat{M}^1|$, because the chance of being correct if classified to $|\hat{M}^0|$ is higher, increasing $FN$ \citep{LeevyClassImbalance}.
We then get a higher denominator for Recall, thus an underestimation of uncertain reference volumes.

Similarly to the Dice, Recall, and Precision do not reward the agreement of emptiness. 
We, therefore, recommend setting a threshold for very small reference volumes and evaluating such cases with image-classification metrics. 

\subsubsection{Distance Agreement}
Overall, we found that distance metrics, especially SDT, show favorable behavior in the context of small and uncertain reference annotations, while still exhibiting a consistent correlation to metrics that measure volume and overlap agreement. 


\textbf{SDT:}

\emph{Robustness toward Uncertainty and Independence from Reference Volume:}\\
SDT assigns cardinalities to surface voxels based on their proximity to the nearest surface voxel in either the reference or predicted mask. 
This approach emulates the behavior of the Dice while serving as a distance metric. However, contrary to Dice, if the reference and predicted volumes are right next to each other and within the border region $\hat{B}^t$, SDT still measures this agreement. 
This becomes particularly advantageous when a lower signal caused by pathophysiological factors and modality-related effects introduces more uncertainty in the outer regions of the target object compared to its inner regions, i.e. like a stroke on NCCT. 
Compared to the Dice, we found weaker correlations to both the U-score and reference volume for the NCCT data set (-0.37, respectively).

In the Spinal cord data set, there is a correlation between SDT to the U-score. Image analysis of a few distinct cases with high uncertainty and low SDT value revealed deteriorating image quality in the cranial and caudal slices of the spinal cord, which is suggested to be the primary source of this relationship. 

\emph{Reward of Volume and Location Agreement:}\\
SDT shares similarities with overlap measures, due to its reliance on the spatial relationships among surface voxels and the direct influence of the object size on $|B^t|$.
Consequently, SDT captures both the agreement in location and volume, which is further supported by its strong correlation with volume and distance metrics across all data sets (0.52-0.87).

\emph{Reward of Agreement of Emptiness:}\\
In the presence of empty reference masks, all distance metrics return "inf". 
Similarly, to overlap metrics, distance metrics may need a lower bound volume threshold to evaluate empty and small volume reference masks with image-classification metrics. 

\textbf{HD 95 and ASSD:}

Given that HD 95 and ASSD are metrics based on distance (as discussed in Section \ref{sec:distance_based}), which implies that if the model predicts a slightly different volume, HD 95 and ASSD should still yield values close to an optimal result, primarily capturing location accuracy and allowing volume error.

Since HD 95 and ASSD are distance-based metrics (as explained in Section \ref{sec:distance_based}),  they primarily assess location accuracy while accommodating for volume errors. Values are close to the optimal even if the model's predicted volume slightly deviates from the reference.

Consistent with this, we found that HD 95 and ASSD exhibited mostly no correlations with reference volumes and uncertainty. 
However, similar to SDT, we observed a correlation between ASSD and the U-score in the Spinal Cord data set, likely attributed to low image quality in the cranial and caudal slices in distinct cases.

Overall, HD 95 and ASSD show robustness to uncertainty and reference volume, however, mostly measure distance agreement. 
Even though they empirically show strong correlations with volume and overlap metrics across all datasets, SDT should be preferred as a metric, if volume and location agreement is crucial. 

\begin{table*}[b]
\caption{Suggestions for Choosing Meaningful Metrics for Data Sets with Uncertain, Small and Empty Reference Annotation \label{tab:guidelines}}
\centering
\begin{tabular}{|l|l|>{\centering\arraybackslash}p{3cm}|>{\centering\arraybackslash}p{3cm}|>{\centering\arraybackslash}p{3cm}|>{\centering\arraybackslash}p{3cm}|}
\hline
\textbf{Category} & \textbf{Metric$^1$}& \textbf{Robustness toward Uncertainty in Reference Annotation} & \textbf{Independence from Volume of Reference Annotation} & \textbf{Reward of Volume and Location Agreement} & \textbf{Reward of Agreement of Emptiness} \\ \hline
\textbf{Volume} & \textbf{VS}     & \checkmark  &  \checkmark   &  -    & \hfil \checkmark \newline   \\ 
    &\textbf{AVD}    & \checkmark   & -   &  -    & \hfil \checkmark \newline  \\ \hline
\textbf{Overlap} & \textbf{Dice}   & -    &  -   &  \checkmark    & \hfil - \newline set~ $threshold$$^2$  \\ 
    &\textbf{Recall} & -   &  -   &  \checkmark    & \hfil - \newline set~ $threshold$$^2$ \\ 
    &\textbf{Precision} & -   &  -   &  \checkmark    & \hfil - \newline set~ $threshold$$^2$ \\ \hline
\textbf{Distance} & \textbf{HD 95}  & \checkmark&  \checkmark  &  -    & \hfil - \newline set~ $threshold$$^2$\\ 
    &\textbf{ASSD}    &  (\checkmark ) & -  &  \checkmark    & \hfil - \newline set~ $threshold$$^2$\\
&\textbf{SDT small}& \checkmark   &  \checkmark   &  \checkmark    & \hfil - \newline set~ $threshold$$^2$ \\ 
&\textbf{SDT large}& \checkmark  &  \checkmark   &  \checkmark    & \hfil - \newline set~ $threshold$$^2$ \\ \hline
\end{tabular}
\footnotesize{\newline $^1$ VS = Volumetric Similarity, AVD = Absolute Volume Difference,ASSD = Average Surface Distance, HD 95 = Hausdorff Distance 95th percentile, SDT = Surface Dice at Tolerance, $^2$ set $threshold$ volume = below this volume threshold images are considered to have no lesion}
\end{table*}
\subsection{Evaluation of Image-level classification Metrics}
\label{sec:image_classification_guidelines}
\todo[inline]{rephrased paragraph, fused results and discussion}
We propose a simultaneous evaluation of image-level classification metrics besides segmentation metrics to ensure an unbias evaluation of model performance when trained on data sets that include cases with uncertain, small, or empty reference annotations.  
Negligible reference volumes below a certain $threshold$ may only be evaluated with image classification metrics

For example, Liu et al. only included positive cases and proposed image classification metric, LDR \citep{LiuDWIstroke}. 
However, the agreement in image-level classification is not assessed in the case of empty reference masks or small-volume cases. Clinical tests are unlikely to exclusively be performed on patients with a present pathology. 

Data sets with negative cases or cases with negligible reference volumes would be more representative of the distribution of patients in clinical practice. 
This can have major implications for the idea of mostly positive or ambiguous cases being read by a radiologist and negative cases confidently evaluated by an algorithm \citep{wang2021ai}.

In this section, we briefly highlight how a class imbalance between positive and negative/small-volume cases also introduces evaluation biases for inter-models and inter-data set comparison.

\begin{figure}
    \centering
    \includegraphics[width=0.48\textwidth]{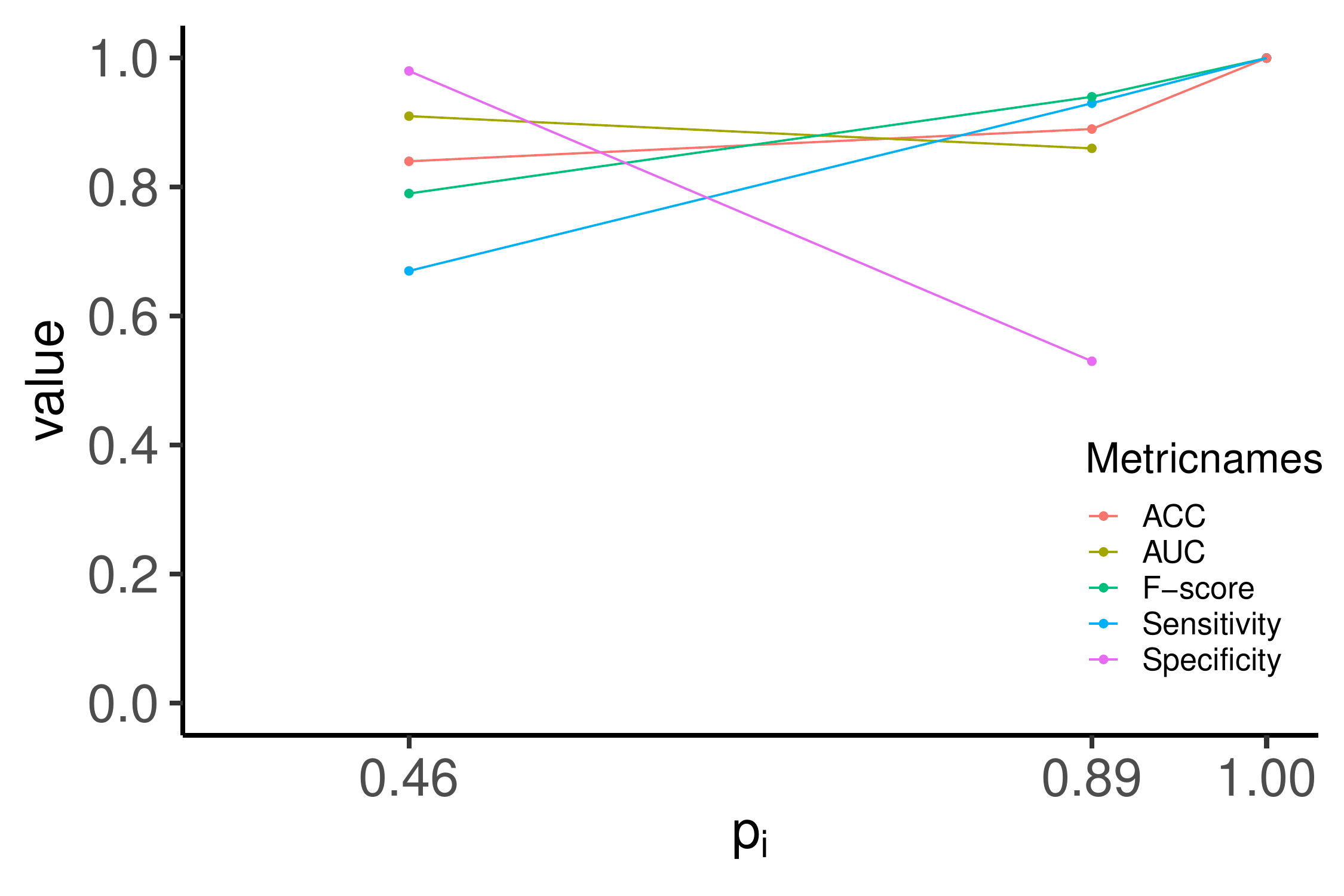}
    \caption{Line plot of image classification metrics value over $p_i$ for the NCCT and BRATS 2019 non-enhancing tumor and whole tumor data set, respectively}
    \label{fig:spaghetti}
\end{figure}

\textbf{Sensitivity, Specificity, and F$_1$-score:}
Whether to use Specificity or Sensitivity as the primary image-classification metric depends on the clinical context. 
For this study, we found higher Sensitivity, Specificity, and F$_1$-score associated with higher $p_i$ (Supplemental material Figure \ref{fig:spaghetti}). However, further studies are needed for more general statements. 

\textbf{ACC: }
The ACC evaluates the agreement in image-level classification in the case of $TN_i$ \emph{and} $TP_i$ \citep{Maier_Hein_pitfall_metric}.
As for Sensitivity, Specificity, and F$_1$-score, we found that a higher ACC value is associated with higher $p_i$ (Supplemental material, Figure \ref{fig:spaghetti}).

\textbf{AUC: }
AUC is a standard multi-threshold classification metric to evaluate a predictor and is not defined for populations where only one class is present (therefore only NCCT and BRATS non-enhancing tumor) \citep{Maier_Hein_pitfall_metric}.
As the true discrete class in this setting is defined by the volume $threshold$, the AUC reveals information on how well the models classify volumes. We found that AUC does not change with $p_i$, suggesting AUC is a more robust metric for unbalanced data sets. 

\section{Limitations}
\todo[inline]{corrected language errors}
The first limitation is that we evaluate metrics behavior and reference annotation uncertainty in only three medical neuroimaging data sets and examine four different target objects.  We introduce methodologies aimed to be applied to a broader range of medical imaging data sets, allowing for a comprehensive examination of our findings.
The second limitation is that the choice of baseline models might influence the correlations between metrics. 
In order to mitigate this, we choose nnUNet, a model that is generalizable to many medical segmentation tasks \citep{IsenseennUNet}.
Furthermore, correlation does not prove causation. 
For example, the correlation between reference volumes and uncertainty to the value of metrics does not imply that a higher reference volume value causes a higher metric value. 
The correlation of Dice and reference volumes have been found in previous works \citep{TahaMetric,LiuDWIstroke,CommowickMSSEG,Maier_Hein_pitfall_metric} however analysis of data sets properties, in-depth analysis, quantification of the uncertainty were missing.
\section{Conclusion}
\todo[inline]{corrected language errors}
We notice a mismatch between dataset properties in challenge-winning segmentation models and cases encountered in clinical practice. 
Some commonly used metrics (i.e. Dice score) might not capture whether models' performance generalize well to the distribution of images encountered in clinical practice. 
In particular, (i) the presence of uncertainty in reference annotations causes misleading values, (ii) small reference volumes lead to unreasonable low metric values, (iii) empty reference annotations cause a return of "NaN", "inf" or zero. 
For a data set with uncertain, small, and empty reference annotations, we suggest that model performance generalizes better to clinical practice when evaluated by the Surface Dice at Tolerance.
We further proposed to set a lower volume threshold for very small volumes or empty reference masks and use image-level classification metrics such as AUC ( \href{https://github.com/SophieOstmeier/UncertainSmallEmpty.git}{USE-Evaluator}).

It is crucial to evaluate the performance of the model using multiple metrics that effectively encompass the specific objectives of the clinical segmentation task. 
These objectives can vary significantly across different areas of clinical practice. 
To facilitate the selection of appropriate metrics, we recommend referring to Table \ref{tab:guidelines}.

We highlight the difficulty of comparing models trained to address different clinical problems.  
While uncertain, small, and empty reference annotations require a rethinking of evaluation, it also increases the value an algorithmic tool provides because the underlying task is hard for human experts.

\section{Acknowledgements}
We would like to thank Georg Schramm for the critical review of this manuscript. 

\bibliographystyle{model2-names.bst}\biboptions{authoryear}
\bibliography{refs}

\section{Supplemental Material}
\FloatBarrier

\subsection{Dice Score of the Random Model \label{app:randomdice}}
We aim to show how the Dice score depends on the volume of the target object, independently of the model performance. We do this by showing that the trend of the Dice score, with respect to volume, is present in the Dice score for a simple, random model. Furthermore, we show that this trend is replicated in multiple settings.

We define the random model for a parameter $IR$, as one where each voxel is chosen to be positive in the predicted mask with a probability $IR$ and there are exactly $IR \times |M^{ROI}|$ voxels in the target object. 
Note that the expected number of predicted positive voxels under this model is exactly $IR \times |M^{ROI}|$.

Under this model, we can compute the expected Dice score, $E_D$, across multiple draws. We use the standard combinations notation, ${{a} \choose {b}}$ to denote the number of orderings where we flip $b$ heads from $a$ coin flips. Note that by definition, $TP + FN = IR  |M^{ROI}|$, the size of the target object.

\begin{align}
\label{eq:DiceOverAlpha}
E_D(p) &= 2^{-|M^{ROI}|}\sum D(TP, TN, FP, FN) Pr[TP, TN, FP, FN] \\
\textrm{where}\notag\\
D(\ldots) &= \left (\frac{2\times TP}{2\times TP + FN + FP} \right) \\
\textrm{and}\notag\\
Pr[\ldots] &= {{TP + FN \choose TP}}{{TN + FP \choose TN}}  \times p^{TP + FP} (1 - p)^{FN + TP}
\end{align}

\FloatBarrier

\begin{table*}
\caption{Reference Annotation Uncertainty: Inter-expert and Majority-expert Agreement}
\label{tab:interexpert}
\begin{tabular}{|l|l|rlrl|rlrl|}
\hline
\textbf{Categories} & \textbf{Metric$^2$} & \multicolumn{2}{l}{\textbf{\begin{tabular}[c]{@{}l@{}}NCCT\\ Inter-expert$^1$\end{tabular}}} & \multicolumn{2}{l|}{\textbf{\begin{tabular}[c]{@{}l@{}}NCCT\\ Majority-expert$^{1}$\end{tabular}}} & \multicolumn{2}{l}{\textbf{\begin{tabular}[c]{@{}l@{}}Spinal\\ Inter-expert$^1$\end{tabular}}} & \multicolumn{2}{l|}{\textbf{\begin{tabular}[c]{@{}l@{}}Spinal\\ Majority-expert$^{1}$\end{tabular}}} \\ \hline
\textbf{Uncertainty} & \textbf{U-score} & 0.87 & ± 0.05 & &  & 0.39 & ±
0.02 & &  \\ \hline
\textbf{Volume} & \textbf{VS} & 0.50 & ± 0.02 & 0.75 & ± 0.03 & 0.93 & ± 0.01 & 0.95 & ± 0.01 \\
\textbf{} & \textbf{AVD {[}ml{]}} & 10.50 & ± 2.11 & 4.25 & ± 0.98 & 0.13 & ± 0.02 & 0.10 & ± 0.02 \\ \hline
\textbf{Overlap} & \textbf{Dice} & 0.39 & ± 0.05 & 0.67 & ± 0.04 & 0.84 & ± 0.01 & 0.91 & ± 0.01 \\
\textbf{} & \textbf{Precision} & 0.39 & ± 0.05 & 0.64 & ± 0.06 & 0.83 & ± 0.02 & 0.95 & ± 0.01 \\
\textbf{} & \textbf{Recall} & 0.41 & ± 0.04 & 0.91 & ± 0.02 & 0.85 & ± 0.02 & 0.88 & ± 0.01 \\ \hline
\textbf{Distance} & \textbf{ASSD} & 4.75 & ± 0.54 & 2.03 & ± 0.23 & 0.11 & ± 0.01 & 0.07 & ± 0.01 \\
\textbf{} & \textbf{HD 95 {[}mm{]}} & 16.73 & ± 2.12 & 9.49 & ± 1.15 & 0.50 & ± 0.14 & 0.48 & ± 0.13 \\
\textbf{} & \textbf{SDT small$^3$} & 0.40 & ± 0.03 & 0.67 & ± 0.02 & 0.85 & ± 0.07 & 0.93 & ± 0.04 \\
\textbf{} & \textbf{SDT large$^4$} & 0.59 & ± 0.05 & 0.83 & ± 0.03 & 0.85 & ± 0.07 & 0.93 & ± 0.04 \\ \hline
\end{tabular}
\footnotesize{\newline $^1$ per case and data set median ± 95\% Confidence Interval (bootstrapped)\newline$^2$ VS = Volumetric Similarity, AVD = Absolute Volume Difference,ASSD} = Average Surface Distance, HD 95 = Hausdorff Distance 95th percentile, SDT = Surface Dice at Tolerances \newline $^3$ Surface Dice at Tolerance with 2mm for NCCT and BRATS 2019 models and 0.05mm for the Spinal Cord model \newline $^4$ Surface Dice at Tolerance with 5mm for NCCT and BRATS 2019 models and 0.1mm for the Spinal Cord 
\end{table*}

\end{document}